\documentclass[journal,twoside,web]{TMI_STYLE/ieeecolor}

\usepackage{TMI_STYLE/tmi}

\usepackage[switch,columnwise]{lineno}

\usepackage{cite}

\usepackage{amsmath,amssymb,amsfonts}
\usepackage{float}
\usepackage{adjustbox}
\usepackage{stfloats}
\usepackage[justification=justified]{caption}
\usepackage{textcomp}
\usepackage[utf8]{inputenc}
\usepackage{longtable}
\usepackage{url}
\usepackage{xr}
\usepackage{multirow}
\usepackage{xcolor}
\usepackage[colorlinks]{hyperref}
\usepackage{nicefrac}       
\usepackage{microtype}      
\usepackage{nicematrix}
\usepackage{booktabs}
\usepackage{multirow}
\usepackage{comment}
\usepackage{pifont}
\newcommand{\cmark}{\ding{51}}%
\newcommand{\xmark}{\ding{55}}%

\hypersetup{
    colorlinks=true,
    linkcolor=blue,
    filecolor=magenta,      
    urlcolor=cyan,
    pdftitle={BolT},
    }

\usepackage{algorithmic}
\usepackage[ruled,vlined]{algorithm2e}
\SetKwInput{KwInput}{Input}                
\SetKwInput{KwOutput}{Output}              
\floatname{algorithm}{Algorithm}

\renewcommand{\arraystretch}{1}

\definecolor{bondiblue}{rgb}{0.0, 0.58, 0.71}
\definecolor{brightcerulean}{rgb}{0.11, 0.62, 0.74}

\usepackage{chngcntr}
\counterwithin*{subsubsection}{subsection}
\renewcommand{\thesubsubsection}{\thesubsection.\alph{subsubsection}}

\newcommand*{\revhl}{\textcolor{black}}

\usepackage{makecell}

\def\BibTeX{{\rm B\kern-.05em{\sc i\kern-.025em b}\kern-.08em
    T\kern-.1667em\lower.7ex\hbox{E}\kern-.125emX}}

\def\BibTeX{{\rm B\kern-.05em{\sc i\kern-.025em b}\kern-.08em
    T\kern-.1667em\lower.7ex\hbox{E}\kern-.125emX}}
\begin{document}



\title{BolT: Fused Window Transformers for fMRI Time Series Analysis}
\author{Hasan A. Bedel, Irmak Sivgin, Onat Dalmaz, Salman UH. Dar and Tolga \c{C}ukur 
\\
\thanks{
This study was supported in part by a TUBITAK BIDEB scholarships awarded to H.A. Bedel and O. Dalmaz, and by a TUBA GEBIP 2015 fellowship, a BAGEP 2017 fellowship, and a TUBITAK 121N029 grant awarded to T. Çukur. \revhl{Funds have been provided by the European Joint Programme Neurodegenerative Disease Research (JPND) 2020 call “Novel imaging and brain stimulation methods and technologies related to Neurodegenerative Diseases” for the Neuripides project ‘Neurofeedback for self-stImulation of the brain as therapy for ParkInson Disease’. The Neuripides project has received funding from the following funding organizations under the aegis of JPND: The Netherlands, The Netherlands Organization for Health Research and Development (ZonMw); Germany, Federal Ministry of Education and Research (BMBF); Czech Republic, Ministry of Education, Youth and Sports (MEYS); France, French National Research Agency (ANR); Canada, Canadian Institutes of Health Research (CIHR); Turkey, Scientific and Technological Research Council of Turkey (TUBITAK).}}
\thanks{H. A. Bedel, I. Sivgin, O. Dalmaz, S. UH. Dar and T. \c{C}ukur are with the Department of Electrical and Electronics Engineering, and the National Magnetic Resonance Research Center (UMRAM), Bilkent University, Ankara, Turkey (e-mails: \{abedel,  cukur\}@ee.bilkent.edu.tr).}
}
\maketitle

\begin{abstract}
Deep-learning models have enabled performance leaps in analysis of high-dimensional functional MRI (fMRI) data. Yet, many previous methods are suboptimally sensitive for contextual representations across diverse time scales. Here, we present BolT, a blood-oxygen-level-dependent transformer model, for analyzing multi-variate fMRI time series. BolT leverages a cascade of transformer encoders equipped with a novel fused window attention mechanism. Encoding is performed on temporally-overlapped windows within the time series to capture local representations. To integrate information temporally, cross-window attention is computed between base tokens in each window and fringe tokens from neighboring windows. To gradually transition from local to global representations, the extent of window overlap and thereby number of fringe tokens are progressively increased across the cascade. Finally, a novel cross-window regularization is employed to align high-level classification features across the time series. Comprehensive experiments on large-scale public datasets demonstrate the superior performance of BolT against state-of-the-art methods. Furthermore, explanatory analyses to identify landmark time points and regions that contribute most significantly to model decisions corroborate prominent neuroscientific findings in the literature. 
\end{abstract}


\begin{IEEEkeywords}
functional MRI, time series, deep learning, transformer, classification, connectivity, explainability.
\end{IEEEkeywords}


\section{Introduction}

Functional MRI (fMRI) measures blood-oxygen-level-dependent (BOLD) responses that reflect changes in metabolic demand consequent to neural activity \cite{hillman2014coupling,rajapakse1998modeling}. Recording BOLD responses at a unique combination of spatio-temporal resolution and coverage, fMRI provides the means to study complex cognitive processes in the human brain \cite{kubicki2003fmri,wang2005support,papma2017effect,mensch2017learning}. On the one hand, task-based fMRI enables researchers to associate stimulus or task variables with multi-variate responses across the brain \cite{li2009review, venkataraman2009exploring,nishimoto2011}. Regions that are co-activated in the presence of a particular variable are taken to be involved in the cortical representation of that variable \cite{simon2004automatized}, and they are considered to be functionally connected \cite{rogers2007assessing}. On the other hand, characteristic multi-variate responses are also eminent in the absence of external stimuli or task, when the subject is merely resting \cite{niu2021modeling,yeo2011organization,van2010intrinsic,hu2006interregional}. In resting-state fMRI, co-activation patterns are typically used to define networks of brain regions, whose functional connectivity (FC) has been associated with various normal and disease states \cite{greicius2008resting,lei2021diagnosis,iraji2015restin,zhang2017hybrid}. Many prior studies have linked behavioral traits and prominent neurological diseases with FC features of BOLD responses \cite{kong2019spatial,rajpoot2015functional,muller2018influences,anderson2013abnormal}. 


Earlier fMRI studies adopted traditional machine learning (ML) to analyze multi-variate brain responses in order to decode task- or disease-related information. Since these ML methods use relatively compact models, feature extraction is typically employed to reduce dimensionality and factor out nuisance variability \cite{mckeown1998independent,svensen2002ica}. A prominent approach first expresses FC features as the temporal correlations of BOLD responses across separate brain regions, and then uses methods such as support vector machines or logistic regression to classify external variables \cite{pereira2009machine,de2008combining,zhang2015resting,wang2019functional}. Later studies have instead adopted deep learning (DL) given its ability to capture complex patterns in high-dimensional data \cite{heinsfeld2018identification,li2020deep,duncan2019biomedical, mlynarski2019deep, kam2019deep}. Various successful deep models have been proposed in the literature based on convolutional \cite{kawahara2017brainnetcnn}, graph \cite{parisot2018disease}, or recurrent architectures \cite{fan2020deep,wang2021graph} that process FC features. Yet, common FC features primarily reflect first-order inter-regional interactions, potentially disregarding higher-order interactions evident in recorded BOLD responses \cite{lahaye2003functional,hu2007nonlinear}. To more directly assess information in fMRI data, several recent studies have instead built classifiers using recurrent models or vanilla transformer models \cite{dvornek2017identifying,nguyen2020attend,malkiel2021pre} on BOLD responses. While powerful, \revhl{these recent architectures can introduce high computational burden when processing long time series,} and they do not embody explicit mechanisms to capture contextual representations of multi-variate data across diverse time scales \cite{ismail2019input,liegeois2019resting,allen2014tracking}.

Here we propose a novel transformer architecture that directly operates on BOLD responses, BolT, for fMRI time-series classification. \revhl{To capture local representations, BolT splits the time series into temporally-overlapping windows and employs a cascade of transformer blocks to encode window-specific representations of BOLD tokens (i.e., linear projections of responses measured across the brain at specific time points)}. To enhance expressiveness across broad time scales without elevating computational costs, BolT leverages a novel fused window attention mechanism that \revhl{utilizes cross attention and token fusion among overlapping windows. While cross attention enables interactions between base BOLD tokens in a given window and fringe tokens in neighboring windows prior to encoding, token fusion enables integration of encoded representations across neighboring windows. To hierarchically transition from local to global representations, the extent of window overlap in transformer blocks is progressively increased across the cascade. BolT improves task performance by utilizing classification ($CLS$) tokens to capture task-oriented high-level features. Window-specific $CLS$ tokens are introduced to maintain local sensitivity and compatibility with the hierarchical model structure. Meanwhile, task-relevant information exchange is promoted by a novel cross-window regularization that aligns these $CLS$ tokens across windows. At the end of the cascade, the encoded $CLS$ tokens are averaged across windows and a linear projection layer is used for classification.}

Comprehensive demonstrations are reported for classification tasks on public datasets: \revhl{gender detection from resting-state fMRI scans and cognitive task detection from task-based fMRI scans in the Human Connectome Project (HCP) dataset \cite{van2013wu}, and disease detection from resting-state fMRI scans in the Autism Brain Imaging Data Exchange (ABIDE) dataset \cite{di2014autism}.} BolT \revhl{achieves higher classification performance than} prior traditional and deep-learning methods, including convolutional, graph, recurrent and transformer baselines. \revhl{Ablation studies are presented that demonstrate the significant contribution of individual design elements to model performance, including learnable $CLS$ tokens, split time windows, token fusion, cross attention, and cross-window regularization.} To interpret the representational information captured by BolT, we devise an explanatory technique on the fused window attention operators. \revhl{The proposed technique extracts gradient-weighted attention maps across the cascade to construct an importance map for BOLD tokens, and thereby identify landmark time points. A logistic regression analysis on the landmark points in then performed to identify brain regions that contribute most significantly to the model's decision.} Explanatory analyses reveal task timings and relevant brain regions that corroborate established neuroscientific findings in the literature. Code for implementing BolT is publicly available at \url{https://github.com/icon-lab/BolT}.

\subsubsection*{\textbf{Contributions}}
\begin{itemize}
    \item We introduce a novel transformer architecture to efficiently and sensitively analyze fMRI BOLD responses. 
    \item A novel fused window attention mechanism is proposed with progressively grown window size to hierarchically capture local-to-global representations.  
    \item A novel cross-window regularization is proposed on global classification features to align high-level representations across the time series. 
    \item An explanatory technique is introduced for BolT that evaluates the relevance of individual time points and brain regions to the classification decisions. 
    \end{itemize}

\section{Related Work}
\subsection{Traditional methods}
Whole-brain fMRI data carry densely overlaid patterns of multi-variate responses, which can be difficult to isolate via uni-variate analysis \cite{penny2011statistical,woolrich2001temporal}. This has sparked interest in adoption of ML for multi-variate fMRI analysis \cite{norman2006beyond,haxby2012multivariate}. Earlier studies in this domain used traditional classifiers such as support vectors machines \cite{song2014svm,wang2007support,hojjati2017predicting}. Because high-dimensional data are paired with models of limited complexity, feature selection is key to improving sensitivity in traditional models \cite{bullmore1996functional,xie2009brain,poldrack2007region}. Accordingly, many traditional models are built on FC features \revhl{derived from response correlations among} brain regions-of-interest (ROIs), as these features are commonly considered to capture discriminative information about cognitive state \cite{zeng2012identifying,shen2010discriminative, khazaee2016application}.

\subsection{Deep learning methods on FC features}
In recent years, DL models have been adopted to elevate sensitivity in fMRI analysis. Some studies have used multi-layer perceptron (MLP) or convolutional neural network (CNN) models to extract high-level features of fMRI data \cite{suk2016state,koyamada2015deep,huang2017modeling} and then to classify external variables \cite{sarraf2016classification,sarraf2016deep,zhao2017automatic}. More commonly, classification models have been built based on FC features among brain ROIs for improved performance \cite{meszlenyi2017resting,kawahara2017brainnetcnn,xing2019dynamic}. Given the brain's intrinsic structure, graph neural networks (GNN) have gained traction wherein individual ROIs denote nodes and FC features among ROIs determine edge weights \cite{li2021braingnn,li2019graph}. To capture temporal variability in dynamic FC features, recurrent or transformer architectures have also been integrated to process the GNN outputs \cite{kim2021learning}. However, GNN-based models might suffer from over smoothing \cite{chen2020measuring} or squashing \cite{alon2020bottleneck} that can lower sensitivity to long-range dependencies. Furthermore, while methods that receive FC features as input can improve learning efficiency by mitigating nuisance variability, FC features typically reflect first-order interactions among ROIs, neglecting potential non-linear effects \cite{su2013discriminative}. 

\subsection{Deep learning methods on BOLD responses}
\textbf{Recurrent networks:} \revhl{An alternative approach to building models on pre-extracted FC features is to directly analyze BOLD responses in fMRI time series}. Given the high degree of temporal correlation in BOLD responses, recurrent neural networks (RNNs) have been proposed to \revhl{sequentially process fMRI data across time} given CNN-based or ROI-extracted spatial representations \cite{li2020detecting,dvornek2017identifying,zhao20203d}. \revhl{Previously reported recurrent architectures in the fMRI literature include vanilla long short-term memory (LSTM) models \cite{dvornek2017identifying}, and hybrid convolutional LSTM models \cite{li2020detecting,zhao20203d}. While recurrent architectures are powerful in time series analysis, sequential processing introduces difficulties in model training on long time series due to vanishing gradients,} and hence they may show suboptimal sensitivity to long-range interactions \cite{kerg2020untangling}. 

\textbf{Vanilla transformers:} \revhl{Transformer architectures based on self-attention mechanisms have recently been introduced to address limitations of recurrent networks \cite{vaswani2017attention}. Given a sequence of tokens, self-attention operators filter their inputs based on inter-token similarity to integrate long-range contextual information. A feed-forward network block, typically selected as an MLP, then encodes latent representations of the contextualized tokens. Several recent studies have employed vanilla transformers that process the entire fMRI time series as a single sequence \cite{nguyen2020attend,zhang2022diffusion}. Vanilla transformers can manifest relatively limited sensitivity to local representations while emphasizing long-range temporal interactions. Moreover, they introduce quadratic computational complexity with respect to sequence length as self-attention requires similarity assessment between all pairs of tokens.}

\textbf{Efficient transformers:} To sensitively analyze fMRI data across diverse time scales while mitigating computational burden, here we introduce a novel transformer architecture, BolT, based on a fused window self-attention mechanism (FW-MSA, Section \ref{FWMSA_sect}). Unlike previous methods that receive as input pre-extracted FC features \cite{abraham2017deriving,parisot2017spectral,gadgil2020spatio,li2021braingnn,kim2021learning}, BolT performs learning on BOLD responses to improve sensitivity. Unlike vanilla transformers that process the time series as a single sequence to focus on global temporal representations \cite{nguyen2020attend,zhang2022diffusion}, BolT improves efficiency by splitting the time series into overlapping windows, and employs a cascaded transformer encoder that hierarchically extracts local-to-global representations. 

\revhl{Several recent studies have devised efficient transformer models with partially similar aims to our proposed approach. A computer vision study has introduced SwinT that restricts self-attention computations to non-overlapping local windows in a given sequence, and performs half-sequence-length shifts in the window position across alternating transformer blocks \cite{liu2021swin}. SwinT implicitly captures cross-window interactions via the window shifts, and it does not utilize high-level $CLS$ tokens. Instead, BolT explicitly captures cross-window interactions by using overlapping windows along with cross attention and token fusion between neighboring windows, and it utilizes dedicated $CLS$ tokens to improve classification. A natural language processing study has proposed Longformer that restricts self-attention to a moving local window centered on each token for encoding the token in a single context, and it uses a global $CLS$ token across the sequence that can degrade sensitivity to local representations \cite{beltagy2020longformer}. In contrast, BolT encodes each token appearing in multiple overlapping windows in multiple contexts and then fuses these encodings, and it uses window-specific $CLS$ tokens that are aligned with cross-window regularization for enhanced sensitivity.} 

\revhl{Efficient transformers have also been adopted in medical image analysis tasks. A hybrid CNN-transformer model, IFT-Net, has been proposed that reduces dimensionality of input data with a convolutional module prior to the transformer \cite{zhao2022ift}. While this approach reduces the sequence length for self-attention computations, it can compromise temporal resolution and sensitivity to local representations in time series analysis. Instead, BolT maintains local sensitivity by preserving temporal dimensionality across the cascade. Another study has introduced HATNet as a hybrid CNN-transformer model where the transformer module sequentially computes intra- and inter-window attention on non-overlapping windows \cite{mehta2022end}. This sequential approach can be suboptimal since inter-window computations are performed on self-attention outputs that ignore cross-window interactions. In contrast, BolT simultaneously computes self- and cross-attention in overlapping windows. A recent fMRI study has introduced a cascaded transformer, TFF, that splits the time series into separate windows to focus on local temporal representations \cite{malkiel2021pre}. TFF processes tokens in separate windows independently across the cascade and naively averages the encoded representations over windows, reducing sensitivity to long-range context. Instead, in each stage of the cascade, BolT uses FW-MSA modules to capture interactions that extend over broad time scales via learning-based fusion of information flow across windows.}

\begin{figure*}[t]
  \centering
\includegraphics[width=0.99\linewidth]{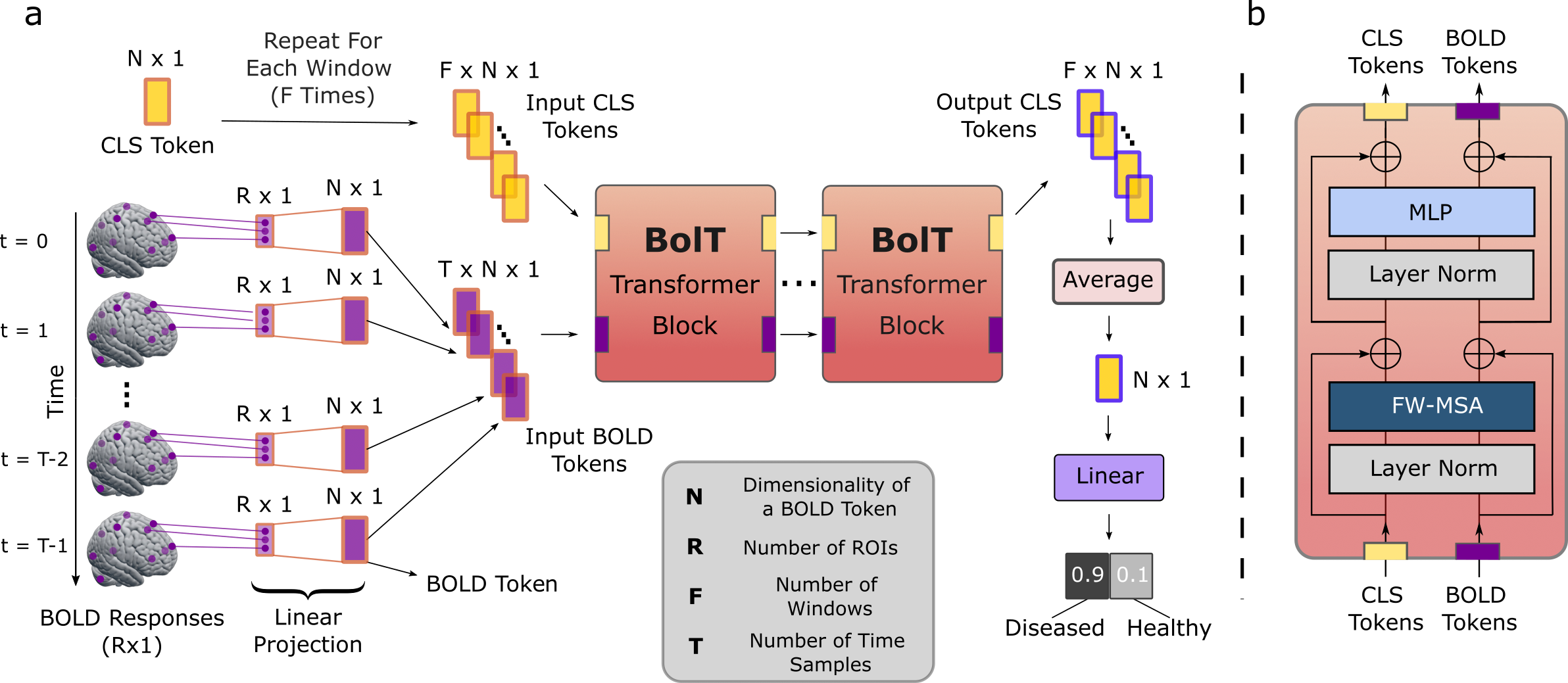}  
  \caption{(a) Overview of BolT. \revhl{First, ROI-level BOLD responses are extracted from four-dimensional fMRI data. These responses are then projected by a learnable linear layer to obtain $T$ BOLD tokens. Each BOLD token encodes ROI responses across the brain recorded at a specific time instant as an $N$-dimensional vector.} A cascade of transformer blocks processes BOLD tokens across a collection of $F$ temporally-overlapping windows within the time series. \revhl{For each time window, a separate learnable classification ($CLS$) token is employed within the transformer blocks. The $CLS$ tokens input to the first block are initialized as tied vectors across $F$ windows, but they become window-specific following encoding through transformer blocks.} The blocks compute latent representations of BOLD and $CLS$ tokens; yet only the CLS tokens are used for the classification task at the output layer. (b) Inner architecture of the transformer block. Unlike vanilla transformers, BolT is equipped with a novel fused window multi-head self-attention (FW-MSA) layer to efficiently capture both local and global context within the fMRI time series.}
\label{fig:BolT}
  
\end{figure*}

\section{Theory}
For multi-variate analysis of four-dimensional (4D) fMRI data recorded in a subject, regional BOLD responses are first extracted  using an external atlas parcellating the brain into $R$ ROIs. The time series for a given ROI is taken as the average response across voxels within the ROI, z-scored to zero mean and unit variance. Our model learns to map these regional BOLD responses $x \in \mathbb{R}^{T\times R}$ (where $T$ is the number of time samples in the fMRI scan) onto class labels $y$ (e.g. subject gender, cognitive task) depending on the task. Note that transformers expect a sequence of tokens as input. Here, we refer to a learnable linear projection of BOLD responses measured at a particular time index as a BOLD token, i.e., $b^{(t)}=f_b(x^{(t)}) \in \mathbb{R}^N$, where $t$ is the time index, $f_b$ is the linear projection, $N$ is the encoding dimensionality. The collection of BOLD tokens across the fMRI scan is then $b = (b^{(0)}, ..., b^{(T-1)}) \in \mathbb{R}^{T\times N}$. Latent representations of BOLD tokens are computed by a cascade of transformer blocks in BolT (Figure \ref{fig:BolT}). The learned latent features $h_f$ are then linearly projected onto individual class probabilities. To capture both local and global representations of BOLD tokens, the transformer blocks split the fMRI time series into ${F}$ overlapping windows, and employ a novel fused window self-attention operator to assess interactions between base tokens in a given time window and fringe tokens in neighboring windows. In this section, we describe the architectural details of the proposed model and introduce an explanatory technique devised for BolT.

\subsection{BolT}
\label{FWMSA_sect}
\begin{figure*}[t]
  \centering
\includegraphics[width=0.975\linewidth]{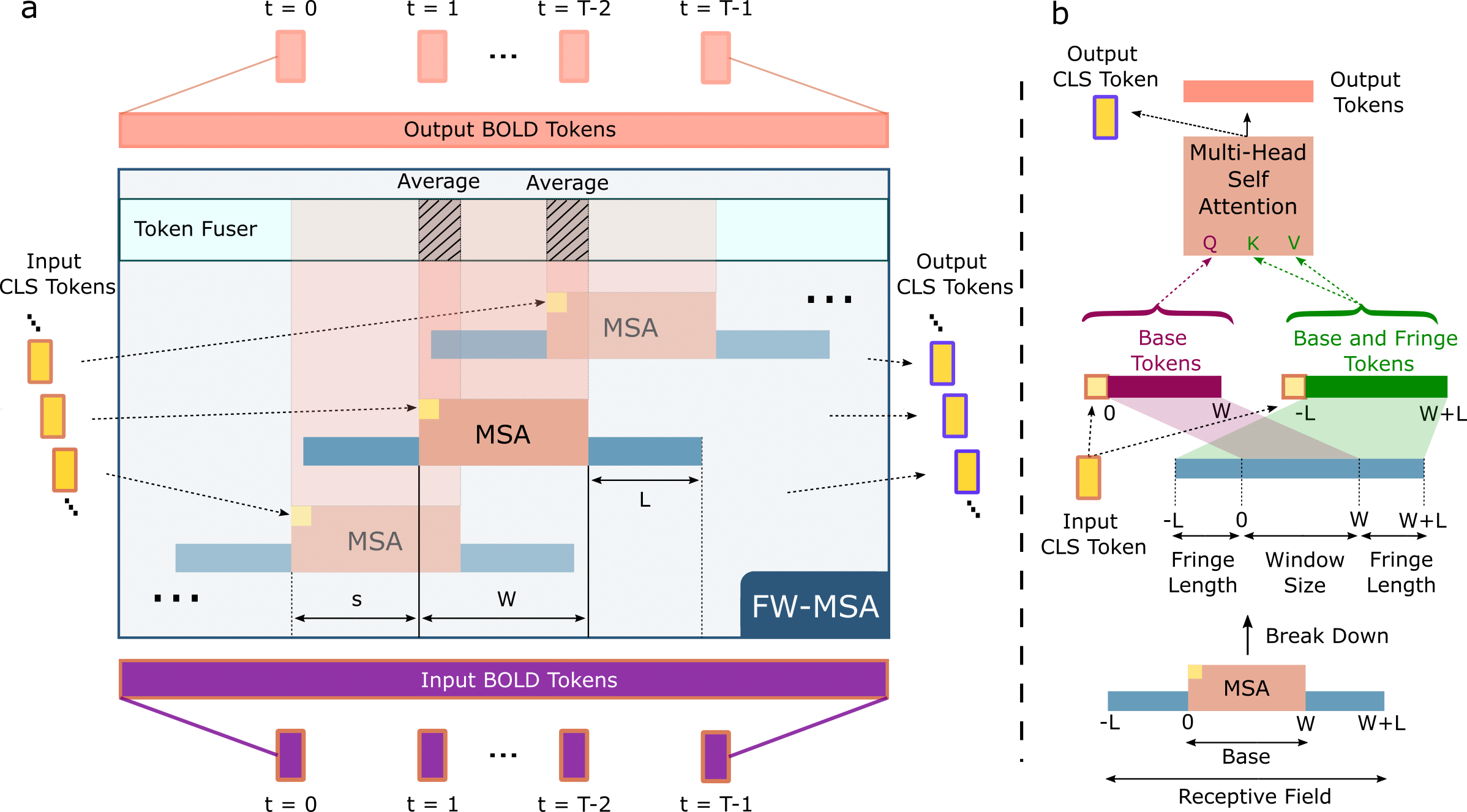}
  \caption{(a) Schematic of the fused window multi-head self-attention (FW-MSA) module. Input BOLD tokens are separated into an overlapping set of time windows of size W and stride s. A CLS token is assigned to each window. A given window possesses a bilateral fringe region of size L to permit interactions with neighboring windows. Within-window interactions are captured via attention among base tokens, whereas cross-window interactions are captured via attention between base and fringe tokens. For each BOLD token, attention-derived latent representations are then fused across the separate time windows in which it appears. (b) Attention calculations within a window. Query to the FW-MSA is a BOLD token from the window base, whereas key and value are BOLD tokens from the broader receptive field including the fringe region.}
    \label{fig:FWMSA}
\end{figure*}

BolT embodies a cascade of transformer blocks and a final linear layer to perform classification. Unlike vanilla transformers \cite{vaswani2017attention}, BolT comprises a novel FW-MSA module to enhance sensitivity to the diverse time scales of dynamic interactions in the brain, while maintaining linear scalability with the duration of fMRI scans (Figure \ref{fig:FWMSA}). A regular MSA layer uses global attention across tokens resulting in quadratic complexity. In contrast, FW-MSA computes local attention within compact time windows extracted from the fMRI scan. Temporal windowing restricts token-to-token interactions to a focal neighborhood surrounding each window. The resultant local precision serves to improve the capture of subtle changes in brain activation dynamics \cite{hutchison2013dynamic}. To capture a window-level latent representation, a ${CLS}$ token is also employed for use in downstream detection tasks \cite{dosovitskiy2020image}. Input ${CLS}$ tokens are initialized as tied vectors across separate time windows. The ${CLS}$ token for each window is concatenated to the query, value and key tokens in FW-MSA. The final layer uses output ${CLS}$ tokens to linearly map their aggregate features onto class logits.

\textbf{Fused window attention:} FW-MSA enables cross-window interactions by attention calculation between base tokens in a given window and fringe tokens in neighboring windows. To do this, FW-MSA first splits the entire collection of BOLD tokens ${b \in \mathbb{R}^{T \times N}}$ into ${F}={(T-W)/s + 1}$ windows of size ${W}$ and stride ${s}$. The receptive field of a given window contains ${W}$ base tokens centrally, and ${L}$ fringe tokens on either side of the base. While processing the $i$-th window, FW-MSA receives as input a collection of ${CLS}$ (${CLS_i \in \mathbb{R}^N}$) and BOLD ($ {b_i \in \mathbb{R}^{(W+2L) \times N}}$) tokens. Let ${Q_i \in \mathbb{R}^{(1+W) \times N}}$ denote queries for base tokens, and ${K_i \in \mathbb{R}^{(1+W+2L)\times N}}$ and ${V_i \in \mathbb{R}^{(1+W+2L)\times N}}$ denote keys and values for the union of base and fringe tokens. Assuming ${f_q, f_k}$ and ${f_v}$ are learnable linear projections, the query, key and value for the ${i}$-th window are:
\begin{eqnarray}
    &&Q_i = f_q( \{ CLS_i, b^{(i \times s)}, ..., b^{(i \times s + W - 1)} \} ) ,\nonumber\\ 
    &&K_i = f_k( \{ CLS_i, b^{(i \times s-L)}, ..., b^{(i \times s + W + L - 1)} \} ), \nonumber\\
    &&V_i = f_v( \{ CLS_i, b^{(i \times s-L)}, ..., b^{(i \times s + W + L - 1)} \} ). 
\end{eqnarray}
To leverage information in the temporal ordering of BOLD tokens, we incorporate a relative position bias in attention calculations \cite{liu2021swin,yang2021focal}:
\begin{gather}
    \mathrm{Attention}(Q_i, K_i, V_i) = \mathrm{Softmax}(\frac{Q_i K_i^T}{\sqrt{d}} + B) V_i,
\end{gather}
where ${B \in \mathbb{R}^{(1+W) \times (1+W+2L)}}$ is a learnable positional bias matrix and $d$ is the feature dimensionality of the attention head. FW-MSA includes multiple attention heads, albeit expressions are given for a single head for simplicity. ${B}$ expresses the positioning of base and ${CLS}$ tokens with respect to all tokens in the receptive field including base, fringe and $CLS$ tokens. For BOLD tokens, $B$ parametrizes the potential range $[-W-L+1, W+L-1]$ of relative distances among tokens in different positions of the receptive field. For the ${CLS}$ token, it instead serves to distinguish the ${CLS}$ token from BOLD tokens.

\textbf{Token fusion:} FW-MSA calculates latent representations of each BOLD token given surrounding local context. Because each token appears in multiple windows, a token fuser is used to aggregate the resultant representations:
\begin{gather}
	b^{(i)}[m] = \frac{1}{P} \sum^{P-1}_{p=0}b^{(i)}_p[m-1],
\end{gather}
where $m \in \{0,1,..., M-1\}$ is the index of the transformer block, $p$ is the index of the time window among $P$ windows that contain a particular token, $b_p^{(i)}[m-1]$ is the $i$-th input BOLD token, and $b^{(i)}[m]$ is the fused token. Token fusion facilitates exchange of information across windows, while maintaining a fixed number of BOLD tokens across transformer blocks. Following fusion, all tokens are forwarded to the MLP module. 

\textbf{Cross-window regularization:} The first transformer block in BolT receives as input a single ${CLS}$ token shared across the time windows. The transformer encoders then compute a unique ${CLS}$ token for each window based on the BOLD tokens within its receptive field. If the latent space of window-level representations captured by the ${CLS}$ tokens are largely incompatible, model performance in downstream classification tasks might be compromised. Thus, to encourage high-level representations that are consistent across time windows, we introduce a novel cross-window regularization as an additional loss term:  
\begin{gather}
    \label{windowConsistencyLoss}
    L_{CWR} = \frac{1}{N F} \mbox{ } \sum_{i=0}^{F-1} || {CLS}_{i}[M-1] - \frac{1}{F} (\sum_{j=0}^{F-1} {CLS}_{j}[M-1]) ||_2^2,
\end{gather}
where ${CLS_{i}[M-1]}$ is the encoded ${CLS}$ token for the $i$-th window at the output of the last transformer block ($M$-th). The regularization term in Eq. \ref{windowConsistencyLoss} penalizes the deviation of individual ${CLS}$ tokens from their mean over windows.

\subsection{Explanatory technique} \label{interpret}
We introduce an explanatory technique for BolT that generates importance weights for BOLD tokens to assess their contribution to a given decision. To do this, we first derive gradient-weighted attention maps as inspired by recent computer vision studies \cite{chefer2021generic}. Yet, we propose an adapted procedure for map calculation in FW-MSA to cope with the overlap and fusion operations across time windows (Alg. \ref{alg:relevancy}). \revhl{In the proposed procedure, gradient-weighted attention maps are calculated separately for each transformer block and each time window}:
\begin{align}
    \Bar{A}_{mi} &= E_h((\nabla {A_{mi}} \odot {A_{mi}})^+). \label{eqgradatt}
\end{align}
In Eq. \ref{eqgradatt}, $A_{mi} \in  \mathbb{R}^{(1+W) \times (1+W+2L)}$ is the attention map \revhl{produced by the FW-MSA layer} at the $m$-th block for the $i$-th window, \revhl{and the first row and column of $A_{mi}$ are reserved for attention values related to the $CLS$ token. The effect of map values onto the model output is characterized via $\nabla A_{mi}$, i.e., the gradient of the loss function with respect to $A_{mi}$}. Meanwhile, $E_h$ denotes the averaging operator across attention heads \revhl{for aggregation}, $\odot$ is the Hadamard product \revhl{to modulate the attention maps with the gradients}, and $^{+}$ denotes rectification \revhl{to prevent negative values}. \revhl{Within each transformer block,} single-window attention maps are then aggregated to form a global attention map across the entire time series, $\Bar{A}_{G}[m] \in \mathbb{R}^{(F+T) \times (F+T)}$ \revhl{where $F$ is the number of windows, $T$ is the number of BOLD tokens}. \revhl{During aggregation, projections of fringe tokens} that appear in multiple windows are averaged across windows. \revhl{Assuming that $t^{'} = F + i \times s$ is the starting index of base tokens in the $i$-th window, $t^{''} = t^{'} - L + p_l$ \revhl{and} $t^{'''} = t^{'} + W + L - p_r$ respectively denote the starting and ending indices of the fringe tokens, where $p_l = max(0, i \times s - L) - (i \times s - L)$ \revhl{and} $p_r = i\times s + W + L - min(T, i\times s + W + L)$ are correction factors to handle windows near the edges of the time series. Receiving as input $\Bar{A}_{mi}$, the projection in the $i$-th window is then expressed as}: 
\begin{eqnarray}
    &&{\Bar{A}_G[m](i,i) = \Bar{A}_{mi}(0,0)} \label{exLabel1} \\ 
    &&{\Bar{A}_G[m](i, t^{''} : t^{'''}) = \Bar{A}_{mi}(0,1+p_l:-p_r)} \label{exLabel2} \\
    &&{\Bar{A}_G[m](t^{'} : t^{'} + W, i) = \Bar{A}_{mi}(1:  ,0)} \label{exLabel3} \\
    &&{\Bar{A}_G[m](t^{'} :  t^{'} + W, t^{''} : t^{'''} ) = \Bar{A}_{mi}(1:, 1+p_l:-p_r)}    \label{exLabel4}\\
    &&{\Bar{A}_G[m] = \Bar{A}_G[m] \oslash A_{norm}} \label{exLabel5}
\end{eqnarray}
Eq. (\ref{exLabel1}) captures self-attention for ${CLS_i}$, Eq. (\ref{exLabel2}) captures attention between ${CLS_i}$ and BOLD tokens within the receptive field of the $i$-th window, Eq. (\ref{exLabel3}) captures attention between base BOLD tokens and the $CLS_i$ token, and Eq. (\ref{exLabel4}) captures attention between based and fringe BOLD tokens. Note that $(a:-b+1)$ selects between the $(a+1)$-th element from the start and the $b$-th element from the end. \revhl{In Eq. (\ref{exLabel5}), $A_{norm} \in \mathbb{R}^{(F+T) \times (F+T)}$ is an occurrence matrix that captures the number of times each token occurs across windows (i.e., all entries for a given token that appears in $n$ windows are set to $n$), and $\oslash$ is Hadamard division used to normalize for repeated token occurrence}.

\revhl{Next, a token-relevance map $Rel[0]$ that represents the influence of each token onto other tokens in the time series is initialized as an identity matrix in $\mathbb{R}^{(F+T) \times (F+T) }$, implying that each token is initially self-relevant. The normalized attention maps are then used to progressively update the token-relevancy map across transformer blocks where $m \in [0, 1,..., M-1]$}: 
\begin{align}
     Rel[m+1] = Rel[m] + \Bar{A}_G[m] Rel[m].   \label{equmulation} 
\end{align}

\begin{algorithm}[t]
\small
\KwInput  {$ \{ \{A_{0(0)},...,A_{0(F-1)}\},...,\{A_{M-1(0)},...,A_{M-1(F-1)}\} \}$: Set of F single-window attention maps from FW-MSA modules across M transformer blocks.} 
\KwOutput{$Rel[M]$: Relevancy map.} 
$Rel[0] = \textbf{I}_{F+T} \quad \emph{Initialize relevancy map} $

\For{$m=0:M-1$} 
{
	$\Bar{A}_G[m] = \textbf{0}_{F+T} \quad \emph{Initialize global attention map} $
	
	\For{$i=0:F-1$}  
	{
	    $\Bar{A}_{mi} \gets E_h((\nabla A_{mi} \odot A_{mi})^+) \; \emph{Weighted attention map}$
	    
	    $t^{'} \gets F + i \times s$

            $p_l \gets max(0, i \times s - L) - (i \times s - L)$

            $p_r \gets i \times s + W + L - min(T, i \times s + W + L)$

            $t^{''} \gets t^{'} - L + p_l$

            $t^{'''} \gets t^{'} + W + L - p_r$            
	    	    
        $\Bar{A}_G[m](i,i) \gets \Bar{A}_{mi}(0,0) \; \emph{CLS to CLS attention}$
        
        $\Bar{A}_G[m](i, t^{''}:t^{'''}) \gets \Bar{A}_{mi}(0,1+p_l:-p_r) \; \emph{CLS to BOLD} $ 

        $\Bar{A}_G[m](t^{'} : t^{'} + W, i) \gets \Bar{A}_{mi}(1:, 0) \; \emph{BOLD to CLS} $

        $ \Bar{A}_G[m](t^{'} : t^{'} + W, t^{''} : t^{'''}) \gets \Bar{A}_{mi}(1 : , 1+p_l:-p_r)$ $ \emph{BOLD to BOLD} $ 
	}
	$\Bar{A}_G[m] \gets \Bar{A}_G[m] \oslash A_{norm} \quad \emph{Normalize for repeats} $ 
	$Rel[m+1] \gets Rel[m] + \Bar{A}_G[m] Rel[m] \quad \emph{Update rel. map} $ 
}
\Return{$Rel[M]$}  
\caption{Calculation of relevancy map}
\label{alg:relevancy}
\end{algorithm}
Following the calculation of the token-relevancy map at the final FW-MSA module, importance weights for input BOLD tokens \revhl{$w_{imp} \in \mathbb{R}^{T}$} are finally derived as:
\begin{gather}
    {w_{imp} = \frac{1}{F} \, \sum_{i=0}^{F-1} Rel[M](i, \;  F : \;)}.  \label{finalImpWeights} 
\end{gather}
Importance weight of a BOLD token for the classification task is taken as the across-window average of relevancy scores between the ${CLS}$ tokens and the given BOLD token.

\section{Methods}

\subsection{Datasets} \label{Dataset}

Demonstrations were performed on fMRI data from the HCP S1200\footnote{\url{https://db.humanconnectome.org}} \cite{van2013wu} and ABIDE I releases\footnote{\url{https://fcon_1000.projects.nitrc.org/indi/abide/}} \cite{di2014autism}. In HCP S1200, resting-state fMRI data (HCP-Rest) were analyzed to predict gender, and task-based fMRI data (HCP-Task) were analyzed to predict cognitive task. In ABIDE I, resting-state fMRI data were analyzed to detect Autism Spectrum Disorder (ASD). Details about datasets are provided below.

\textbf{HCP-Rest:} Preprocessed fMRI data from 1200 subjects released by the WU-Minn HCP consortium were analyzed \cite{glasser2013minimal}. For each subject, the first session of resting-state scans was used, and incomplete scans with shorter than 1200 time samples were excluded. HCP-Rest comprised a total of 1093 scans from 594 female and 499 male subjects.

\textbf{HCP-Task:} Preprocessed fMRI data from 1200 subjects released by the WU-Minn HCP consortium were analyzed \cite{glasser2013minimal}. The first session of task-based scans was used, where each subject performed seven different tasks in separate runs: emotion, relational, gambling, language, social, motor, and working memory. Incomplete scans were excluded. HCP-Task comprised a total of 7450 scans from 594 female and 501 male subjects. 

\textbf{ABIDE-I:} Preprocessed fMRI data released by the Preprocessed Connectomes Project were analyzed  \cite{craddock2013neuro,di2014autism}. Low-quality scans that did not pass quality checks from all raters were excluded from analysis. ABIDE-I comprised a total of 871 scans from 403 patients with ASD and 468 healthy controls \cite{abraham2017deriving}.

\subsection{Experimental procedures} \label{ExperimentSettings}

Experiments were conducted in PyTorch on an NVIDIA RTX 3090 GPU. Modeling was performed via a nested cross-validation procedure with 10 outer and 1 inner folds. Accordingly, subjects were split into non-overlapping training (80\%), validation (10\%), and test sets (10\%). For fair comparison, all competing methods used identical data splits. For each competing method, hyperparameter selection was performed based on performance in the first validation set and selected parameters that showed near optimal performance across all datasets and atlases were used thereafter. The selected parameters included learning rate $\in(10^{-6},10^{-1})$, number of epochs $\in(5,100)$ and mini-batch size $\in(1,100)$. Training was performed via the Adam optimizer. BolT was trained to minimize the following loss: ${L = L_{CE} + \lambda \cdot L_{CWR}}$ where ${L_{CE}}$ is cross-entropy loss, and ${\lambda}=0.1$ is the regularization coefficient for CWR loss set via cross-validation. Mean and standard deviation of model performance were reported across the test sets. 

For each subject in the training set, fMRI time series were randomly cropped in the temporal dimension to 600 samples for HCP-Rest, 150 samples for HCP-Task, and 100 samples for ABIDE-I to improve stochasticity and learning efficiency \cite{kim2021learning}. Functional data were registered to corresponding structural data for each subject, and aligned to the MNI template. ROI definitions were implemented using two public brain atlases: the Schaefer atlas \cite{schaefer2018local} with 400 regions labeled across seven intrinsic connectivity networks, and the AAL atlas \cite{tzourio2002automated} with 116 regions.

\subsection{Implementation details} \label{configuration}
In this section, the architectural and hyperparameter details of BolT are summarized. BolT was trained for 20 epochs with a batch size of 32 and an initial learning rate of $10^{-4}$. The learning rate was increased to 2x$10^{-4}$ in the first 10 epochs and then gradually decreased to $10^{-5}$. A linear projection layer matched the dimensionality of input BOLD responses to the hidden dimensionality of the transformer blocks. A cascade of four blocks was used, each composed of FW-MSA and MLP modules that used layer normalization and skip connections. A hidden dimensionality of 400 and 40 attention heads with 20 dimensions per head were prescribed. A dropout rate of 0.1 was used in both FW-MSA and MLP layers. For FW-MSA modules, given a desired window size ${W}$, stride ${s}$ and fringe length $L$ were set proportionately as follows:
\begin{gather}
    s = W \alpha, \; \; L = m \,  (W-s)\,\beta  = m\, (1-\alpha)\, W \, \beta,   \label{config}
\end{gather}
where ${m} \in \{0,1, ..., M-1\} $ is the block index, ${\alpha\in \mathbb{R}^+}$ is the stride coefficient (i.e., proportionality constant), and ${\beta\in Z^+}$ is the fringe coefficient. Note that the fringe length was progressively grown over transformer blocks as the number of fused tokens increased. Here, cross-validated search for $W\in(10, 200)$, $\alpha\in(0.0, 1.0)$, $\beta\in(0, 3)$ was performed. Hyperparameters were selected as $W=20$, $\alpha=0.4$, $\beta=2$.

\begin{table*}[!htb]
  \caption{Performance of BolT variants ablated of essential design elements. Ablated elements were learnable $CLS$ tokens ($CLS$), split time windows (Windowing), token fusion (Fusion), cross-attention between the base and fringe tokens (Cross Attn.), and cross-window regularization (CWR). \revhl{When windowing is enabled, the annotation (G) denotes utilization of a global $CLS$ token shared across windows, whereas (L) denotes utilization of local $CLS$ tokens in each window.} Results are shown based on both the Schaefer and AAL atlases. Accuracy, recall, precision and AUC metrics are reported as mean(std) across test folds. Bold-face indicates the top-performing model.}
  \label{tab:ablation}
  \centering
  \resizebox{0.725\textwidth}{!}{%
  \renewcommand{\arraystretch}{0.01} 
    \begin{tabular}{ccccccccccc}
        \toprule
        \thead{Atlas} & \thead{$CLS$} & \thead{Windowing} & \thead{Fusion} & \thead{Cross Attn.} & \thead{CWR} & \thead{Acc. (\%)} & \thead{Rec. (\%)} & \thead{Prec. (\%)} & \thead{AUC (\%)} \\
        \multirow{7}{*}[1em]{\thead{Schaefer}} 
         & \centering \xmark & \centering \xmark & \centering \xmark & \centering \xmark & \centering \xmark & \thead{86.35 \\ $\pm$3.56} & \thead{85.76\\ $\pm$4.94} & \thead{84.74\\ $\pm$4.68} & \thead{94.59\\ $\pm$1.57} \\
         & \centering \cmark & \centering \xmark & \centering \xmark & \centering \xmark & \centering \xmark & \thead{89.65\\ $\pm$3.36} & \thead{88.38\\ $\pm$5.63} & \thead{89.15\\ $\pm$4.78} & \thead{96.03\\ $\pm$1.23} \\
         & \centering {\cmark (G)} & \centering {\cmark} & \centering {\xmark} & \centering {\xmark} & \centering {\xmark} & {\thead{84.99\\ $\pm$2.37}} & {\thead{89.78\\ $\pm$4.76}} & {\thead{79.95\\ $\pm$2.68}} & {\thead{93.63\\ $\pm$1.58}} \\         
         & \centering \cmark {(L)} & \centering \cmark & \centering \xmark & \centering \xmark & \centering \xmark & \thead{89.65\\ $\pm$1.85} & \thead{88.77\\ $\pm$3.27} & \thead{88.71\\ $\pm$3.06} & \thead{96.79\\ $\pm$0.90} \\
        & \centering \cmark {(L)} & \centering \cmark & \centering \cmark & \centering \xmark & \centering \xmark & \thead{90.29\\ $\pm$1.77} & \thead{89.57\\ $\pm$2.54} & \thead{89.34\\ $\pm$3.23} & \thead{97.15\\ $\pm$1.01} \\
         & \centering \cmark {(L)} & \centering \cmark & \centering \cmark & \centering \cmark & \centering \xmark & \thead{91.03\\ $\pm$2.12} & \thead{89.97\\ $\pm$2.70} & \thead{90.42\\ $\pm$3.30} & \thead{97.09\\ $\pm$1.10} \\
         & \centering \cmark {(L)} & \centering \cmark & \centering \cmark & \centering \cmark & \centering \cmark & \thead{\textbf{91.85}\\ \textbf{$\pm$3.05}} & \thead{\textbf{90.58} \\ \textbf{$\pm$4.97}} & \thead{\textbf{91.51}\\ \textbf{$\pm$3.07}} & \thead{\textbf{97.35}\\ \textbf{$\pm$1.06}} \\
        \midrule
        \multirow{6}{*}[1em]{\thead{AAL}} 
        & \centering \xmark & \centering \xmark & \centering \xmark & \centering \xmark & \centering \xmark & \thead{80.01\\ $\pm$3.06} & \thead{78.20\\ $\pm$5.61} & \thead{78.22\\ $\pm$3.88} & \thead{88.74\\ $\pm$3.14} \\
        & \centering \cmark & \centering \xmark & \centering \xmark & \centering \xmark & \centering \xmark & \thead{83.12\\ $\pm$3.58} & \thead{80.80\\ $\pm$5.38} & \thead{82.25\\ $\pm$5.05} & \thead{90.88\\ $\pm$1.91} \\
         & \centering {\cmark (G)} & \centering {\cmark} & \centering {\xmark} & \centering {\xmark} & \centering {\xmark} & {\thead{78.74\\ $\pm$4.11}} & {\thead{85.00\\ $\pm$5.07}} & {\thead{73.02\\ $\pm$4.41}} & {\thead{88.41\\ $\pm$3.41}} \\        
         & \centering \cmark {(L)} & \centering \cmark & \centering \xmark & \centering \xmark & \centering \xmark & \thead{86.85\\ $\pm$3.06} & \thead{87.20\\ $\pm$4.30} & \thead{84.78\\ $\pm$4.65} & \thead{93.57\\ $\pm$2.25} \\
         & \centering \cmark {(L)} & \centering \cmark & \centering \cmark & \centering \xmark & \centering \xmark & \thead{87.04\\ $\pm$2.48} & \thead{87.60\\ $\pm$4.96} & \thead{84.73\\ $\pm$3.43} & \thead{93.80\\ $\pm$2.13} \\
         & \centering \cmark {(L)} & \centering \cmark & \centering \cmark & \centering \cmark & \centering \xmark & \thead{87.13\\ $\pm$3.03} & \thead{85.80\\ $\pm$5.39} & \thead{86.17\\ $\pm$4.22} & \thead{94.12\\ $\pm$1.98} \\
        & \centering \cmark {(L)} & \centering \cmark & \centering \cmark & \centering \cmark & \centering \cmark & \thead{\textbf{87.31}\\ \textbf{$\pm$2.69}} & \thead{\textbf{86.99}\\ \textbf{$\pm$4.49}} & \thead{\textbf{85.65}\\ \textbf{$\pm$4.01}} & \thead{\textbf{94.29}\\ \textbf{$\pm$2.05}} \\               
        \bottomrule
    \end{tabular}}
    
\end{table*}

\begin{table*} [t]
    \caption{\revhl{Performance of BolT under varying window sizes $W$. Results are shown based on the Schaefer and AAL atlases. Accuracy, recall, precision and AUC metrics are reported as mean$\pm$std across test folds. Bold-face indicates the top-performing model.}}
    \label{tab:windowAblation}
    \centering
    \resizebox{0.55\textwidth}{!}{
    \begin{tabular}{ccccccc}
         \toprule
         \thead{Atlas} & \thead{Window size} & \thead{Acc. (\%)} & \thead{Rec. (\%)} & \thead{Prec. (\%)} & \thead{AUC (\%)} \\
         \multirow{4}{*}[1.75em]{\thead{Schaefer}} & \thead{W=10} & \thead{91.39$\pm$2.67} & \thead{89.98$\pm$4.28} & \thead{91.09$\pm$2.81} & \thead{97.49$\pm$0.78} \\
         & \thead{W=20} & \thead{\textbf{91.85$\pm$3.05}} & \thead{\textbf{90.58$\pm$4.97}} & \thead{\textbf{91.51$\pm$3.07}} & \thead{97.35$\pm$1.05} \\
         & \thead{W=80} & \thead{91.48$\pm$2.17} & \thead{90.18$\pm$4.41} & \thead{91.18$\pm$2.86} & \thead{\textbf{97.56$\pm$0.95}} \\  
         & \thead{W=200} & \thead{88.74$\pm$2.73} & \thead{84.77$\pm$5.92} & \thead{90.04$\pm$2.34} & \thead{96.32$\pm$1.44} \\
         \toprule
         \multirow{4}{*}[1.75em]{\thead{AAL}} & \thead{W=10} & \thead{86.58$\pm$2.68} & \thead{86.20$\pm$4.77} & \thead{84.88$\pm$3.98} & \thead{93.83$\pm$2.10} \\
         & \thead{W=20} & \thead{\textbf{87.31$\pm$2.69}} & \thead{\textbf{86.99$\pm$4.49}} & \thead{\textbf{85.65$\pm$4.01}} & \thead{\textbf{94.29$\pm$2.05}} \\
         & \thead{W=80} & \thead{86.39$\pm$4.42} & \thead{86.00$\pm$4.56} & \thead{84.71$\pm$5.88} & \thead{94.17$\pm$2.30} \\
        & \thead{W=200} & \thead{85.57$\pm$4.04} & \thead{82.80$\pm$5.60} & \thead{85.40$\pm$5.31} & \thead{91.98$\pm$3.72} \\
         \bottomrule
    \end{tabular}
    }
\end{table*}


\subsection{Model complexity} \label{complexity}
\revhl{Complexity of models that directly analyze BOLD responses as a temporal sequence depends on the extent of computations performed on the input sequence of BOLD tokens. Assume that the latent dimensionality of tokens is $N$ in a sequence of length $T$. Recurrent architectures process the tokens in the input sequence serially. As such, a recurrent layer follows a well-known ${O(N^2 T)}$ complexity that scales linearly with the sequence length \cite{dvornek2017identifying,zhao20203d,xing2019dynamic}. A convolution layer also has a linear ${O(k N^2 T)}$ complexity where $k$ denotes the size of the convolution kernel. Meanwhile, vanilla transformers use regular MSA layers that exhaustively compute interactions between all time points in the input sequence, with $T$ queries and $T$ keys of dimensionality $N$. Thus, vanilla transformers such as BAnD and IFT-Net incur ${O(NT^2)}$ complexity that scales quadratically with the input sequence length \cite{nguyen2020attend,zhao2022ift}.} 

In contrast, the FW-MSA layer in BolT computes focal interactions between time points in overlapping time windows. Each window has ${W+1}$ queries and ${W+2L+1}$ keys. The complexity within a single window is ${O(NW^2 + NWL)}$. Given a total of ${(T-W)/s}$ windows for the entire sequence, the overall complexity is ${O(NTW^2/s + NTWL/s)}$. \revhl{Selecting ${s}$ and ${L}$ as outlined in Eq.\ref{config}, BolT incurs ${O(NTW\frac{(1+\beta\,(m) \, (1-\alpha) )}{\alpha})}$ complexity that linearly scales with sequence length. Other efficient transformers also show a similar linear trend  \cite{liu2021swin,malkiel2021pre,beltagy2020longformer,mehta2022end}. For instance, $L=0$ (no cross-window attention) and $s=\frac{W}{2}$ result in a linear $O(NTW)$ complexity in SwinT \cite{liu2021swin}, and $L=0$ and $s=W$ result in $O(NTW)$ complexity in TFF and HATNet \cite{malkiel2021pre}.}

\subsection{Competing methods}
BolT was demonstrated against several state-of-the-art methods for fMRI classification including recent transformer, graph, convolutional and recurrent network models, along with a traditional classifier. The architecture, loss function and learning rate scheduler for each competing method were adopted from the original proposing papers. 

\textbf{SVM:}
A traditional model operating on static FC features was considered \cite{abraham2017deriving}. An $\ell_2$ regularized model with linear kernel was used. FC features were computed via Pearson's correlation between ROI-level responses. The cross-validated hyperparameter was a regularization weight of $C$=1.

\textbf{BrainNetCNN:}
A CNN model operating on static FC features of fMRI data was considered \cite{kawahara2017brainnetcnn}. ROI-level features were processed with a cascade of two edge-to-edge, one edge-to-node, and one node-to-graph convolutional layers followed by three linear layers. FC features were computed via Pearson's correlation. Cross-validated hyperparameters were $10^{-4}$ learning rate, 20 epochs, and 16 batch size.

\textbf{BrainGNN:}
A GNN operating on static FC features of fMRI data was considered \cite{li2021braingnn}. Taking ROIs as graph nodes, BrainGNN used a cascade of graph convolutional layers to assign nodes to clusters with learned embeddings, and used pooling layers to aggregate information with element-wise score normalization. FC features were computed via partial correlation \cite{li2021braingnn}. Cross-validated hyperparameters were $10^{-2}$ learning rate, 50 epochs, and 100 batch size.

\textbf{STAGIN:}
A GNN operating on dynamic FC features of fMRI data was considered \cite{kim2021learning}. STAGIN processed shifted windows across the time series by a four-layer GNN taking ROIs as nodes, deriving edges based on FC features, and calculating node features via a recurrent unit. Features extracted from each graph by a squeeze-excitation readout module were consolidated onto a single latent feature by a transformer. The latent feature was linearly projected to class logits. A window size of 50 was prescribed, FC features were computed via Pearson's correlation \cite{kim2021learning}. Cross-validated hyperparameters were 2$\times 10^{-4}$ learning rate, 40 epochs, 8 batch size.


\textbf{LSTM:}
An RNN model operating on BOLD responses was considered \cite{dvornek2017identifying}. A single-layer LSTM model averaged hidden states across time samples, and performed classification via a sigmoid activation layer. Cross-validated hyperparameters were $10^{-3}$ learning rate, 30 epochs, 64 batch size.

\revhl{\textbf{CNN-LSTM:}
A convolutional RNN model operating on BOLD responses was considered \cite{zhao20203d}. CNN-LSTM was adopted to use 1D convolutions. Hidden states were averaged across time samples to perform classification via a sigmoid activation layer. Cross-validated hyperparameters were $10^{-3}$ learning rate, 50 epochs, and 64 batch size.
}

\revhl{\textbf{GC-LSTM:} 
A graph convolutional RNN model operating on FC features was considered \cite{xing2019dynamic}. GC-LSTM was adopted to use the FC features to construct graph adjacency matrices on windowed time series, and it computed recurrent state updates via spectral graph convolution. The window size and stride were set to 120 and 2 sec, respectively \cite{xing2019dynamic}. Hidden states were averaged across time samples to perform classification via an MLP followed by a sigmoid activation layer \cite{xing2019dynamic}. Cross-validated hyperparameters were a learning rate of $10^{-3}$, 100 epochs, and a batch size of 16.
}

\revhl{\textbf{SwinT:}
A transformer model with a windowed attention mechanism proposed for computer vision tasks was considered \cite{liu2021swin}. For fair comparison, the number and hidden dimensionality of transformer blocks, number of attention heads, and window size were matched with BolT. ROI-level responses were linearly projected to match the hidden dimensionality of the transformer blocks. Output tokens were averaged across windows and linearly projected onto class logits for classification \cite{beltagy2020longformer}. Cross-validated hyperparameters were $2 \times 10^{-4}$ learning rate, 30 epochs, and 32 batch size.}

\revhl{\textbf{Longformer:}
A transformer model with a windowed attention mechanism proposed for language tasks was considered \cite{beltagy2020longformer}. Longformer uses sliding window attention for BOLD tokens, and global attention for the $CLS$ token. For fair comparison, the number and hidden dimensionality of transformer blocks, number of attention heads, and window size were matched with BolT. ROI-level responses were linearly projected to match the hidden dimensionality of the transformer blocks. The global $CLS$ token output was linearly projected onto class logits for classification \cite{beltagy2020longformer}. Cross-validated hyperparameters were $2 \times 10^{-4}$ learning rate, 40 epochs, and 32 batch size.}

\textbf{BAnD:} A hybrid CNN-transformer model operating on BOLD responses was considered \cite{nguyen2020attend}. The CNN module suffered from suboptimal learning, so for fair comparison BAnD was implemented with ROI-level inputs as in BolT. ROI-level responses were linearly projected to match the hidden dimensionality of the transformer. No windowing was performed on the time series. Cross-validated hyperparameters were $10^{-4}$ learning rate, 30 epochs, and 32 batch size.

\textbf{TFF:} A hybrid CNN-transformer model operating on BOLD responses was considered \cite{malkiel2021pre}. The CNN module was observed to suffer from suboptimal learning, so for fair comparison TFF was implemented with ROI-level inputs as in BolT. ROI-level responses were linearly projected to match the hidden dimensionality of the transformer. The time series was split into windows of size 20 with shifts of 10, which were processed independently until the final layer where they were averaged \cite{malkiel2021pre}. Cross-validated hyperparameters were $10^{-5}$ learning rate, 30 epochs, and 32 batch size.

\revhl{
\textbf{IFT-Net:} A hybrid CNN-transformer model proposed for medical image analysis was considered \cite{zhao2022ift}. IFT-Net was adopted to use 1D convolutional projections in attention layers where ROI-level responses were taken as input channels. ROI-level responses were linearly projected to match the hidden dimensionality of the transformer blocks. The output feature vector was fed to a sigmoid activation layer for classification. Cross-validated hyperparameters were $10^{-4}$ learning rate, 30 epochs, and 8 batch size.
}

\revhl{\textbf{HATNet:}
A hybrid CNN-transformer model proposed for medical image analysis was considered \cite{mehta2022end}. HATNet was adopted to use 1D CNN modules with ROI-level responses taken as input channels. The output of the CNN module was split into windows of size 16 and processed via the transformer module for classification \cite{mehta2022end}. Cross-validated hyperparameters were $10^{-4}$ learning rate, 100 epochs, and 32 batch size.}

\begin{figure*}[t]
  \centering
\includegraphics[width=0.85\linewidth]{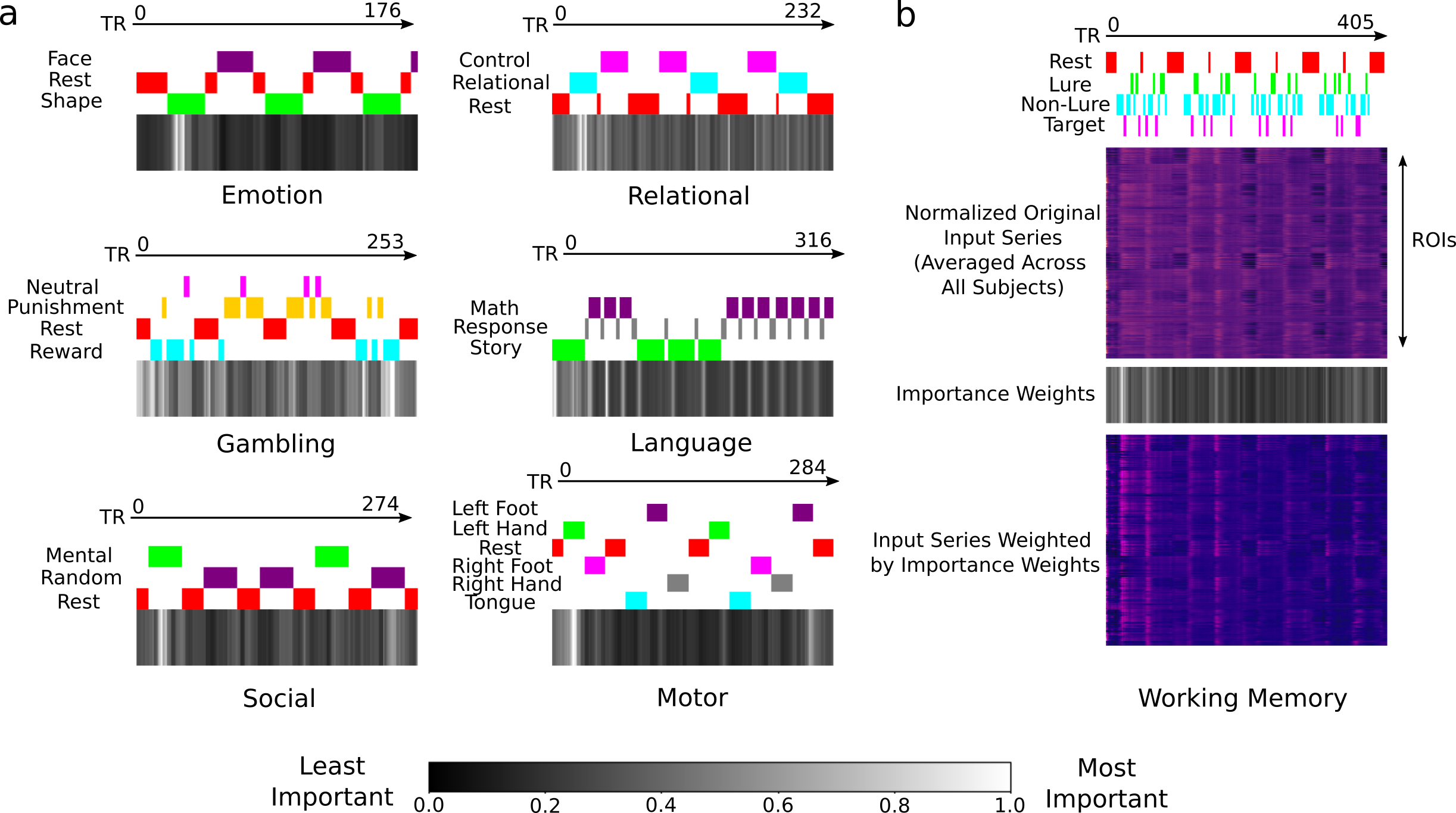}  
    \caption{Importance weights for individual cognitive tasks in HCP-Task averaged across subjects (see colorbar). (a) Results for emotion, relational, gambling, language, social and motor tasks. Subtask annotations are shown to outline task structure during fMRI scans. (b) Input fMRI time series and importance weights for the working memory task. Relevant ROIs and time points are emphasized by masking the input time series with importance weights.}
    \label{fig:relevancyFigure}
\end{figure*}

\begin{table}[t] 
  \caption{Performance of logistic-regression models given as input most important tokens determined via the explanatory technique. Results are also given for logistic-regression models based on random tokens. Boldface indicates the top-performing model in each classification task. }
  \label{logisticTable}
  \centering
  \resizebox{0.625\columnwidth}{!}{%
    \renewcommand{\arraystretch}{0.01} 

  \begin{tabular}{lll}
    \toprule
 & 
    \thead{LR (Important)} &
    \thead{LR (Random)} \\

    \midrule
    \thead{HCP-Rest} & & \\
     \thead{Acc. (\%)} & \thead{\textbf{77.74}\textbf{$\pm$6.82}} & \thead{49.59$\pm$2.79} \\
     \thead{Rec. (\%)} & \thead{\textbf{71.93}\textbf{$\pm$6.43}} & \thead{26.86$\pm$3.76} \\
     \thead{Prec. (\%)} & \thead{\textbf{77.98}\textbf{$\pm$8.47}} & \thead{41.98$\pm$4.37} \\
     \thead{AUC (\%)} & \thead{\textbf{83.70}\textbf{$\pm$5.64}} & \thead{45.70$\pm$1.25} \\
    \midrule
    \thead{HCP-Task} & & \\
     \thead{Acc. (\%)} & \thead{\textbf{94.83}\textbf{$\pm$1.44}} & \thead{15.14$\pm$1.18} \\
     \thead{Rec. (\%)} & \thead{\textbf{94.81}\textbf{$\pm$1.44}} & \thead{15.15$\pm$1.18} \\
     \thead{Prec. (\%)} & \thead{\textbf{94.89}\textbf{$\pm$1.43}} & \thead{15.51$\pm$1.47}  \\
     \thead{AUC (\%)} & \thead{\textbf{99.22}\textbf{$\pm$0.33}} & \thead{51.19$\pm$0.73} \\
    \midrule
    \thead{ABIDE-I} & & \\
     \thead{Acc. (\%)} & \thead{\textbf{60.44}\textbf{$\pm$3.55}} & \thead{51.23$\pm$4.05} \\
     \thead{Rec. (\%)} & \thead{\textbf{46.81}\textbf{$\pm$4.80}} & \thead{33.11$\pm$5.36} \\
     \thead{Prec. (\%)} & \thead{\textbf{59.71}\textbf{$\pm$4.84}} & \thead{47.68$\pm$6.31} \\
     \thead{AUC (\%)} & \thead{\textbf{62.26}\textbf{$\pm$4.65}} & \thead{49.33$\pm$5.34}  \\
    \bottomrule
  \end{tabular}}
\end{table}

\section{Results}
 \label{results}

\subsection{Ablation studies}

\revhl{We conducted a series of ablation studies to evaluate the contribution of individual design elements in BolT. The design elements included learnable and local (i.e., window-specific) $CLS$ tokens, split time windows, token fusion, cross attention between base and fringe tokens, and cross-window regularization. Starting with a vanilla transformer variant, ablated variants were obtained by progressively introducing individual elements. The vanilla variant omitted all elements including $CLS$ tokens, so classification was performed by linearly projecting the time average of encoded BOLD tokens at the output of the last transformer block. For all variants, the architecture and hyperparameters for the utilized components were matched with BolT. To assess the contribution of a learnable $CLS$ token, a variant was formed by introducing a global $CLS$ token into the vanilla variant. Note that local $CLS$ tokens are not applicable in this case since windowing was omitted. To assess the contribution of windowing, two variants were formed that used split time windows and either global or local $CLS$ tokens, albeit omitted token fusion, cross attention and cross-window regularization. Since cross attention was not used, these variants split the time series into non-overlapping windows by setting stride equal to window size and fringe coefficient to 0 (i.e., $s=W$, $L=0$). For the global $CLS$ variant, attention for the $CLS$ token was computed with all BOLD tokens in the time series, whereas attention for a BOLD token was restricted to the window it resided in. For the local $CLS$ variant, attention computations for all tokens were restricted to local windows. To assess the contribution of token fusion, a variant was formed that used local $CLS$ tokens, splitting into overlapping windows (i.e., $s=W \alpha$ as in BolT) and token fusion, albeit omitted cross attention and cross-window regularization. Cross attention was omitted by setting $L=0$. To assess the contribution of cross attention, a variant was formed that used local $CLS$ tokens, splitting into overlapping windows, token fusion and cross-window attention (i.e., $L=m(1-\alpha)W \beta $), albeit omitted cross-window regularization. Finally, to assess the contribution of cross-window regularization, a variant using all elements (i.e., BolT) was employed where regularization was achieved via adding the loss term in Eq. \ref{windowConsistencyLoss}.}

\revhl{Table \ref{tab:ablation} lists performance metrics for all ablated variants. First, we find that introduction of a learnable, global $CLS$ token into the vanilla variant consistently improves performance metrics, demonstrating the utility of high-level $CLS$ over low-level BOLD tokens in classification tasks. Second, a performance loss is incurred when the global $CLS$ token is used along with split time windows, suggesting that the global token does not adequately capture local representations. In contrast, local $CLS$ tokens in combination with windowing yield a performance boost, indicating the importance of using window-specific $CLS$ tokens to learn local representations. Third, fusion of tokens with repeated occurrence in overlapping windows yields a further improvement. Note that the BOLD token for a given time point appears in a single window and is encoded in a single context for a non-overlapping split, whereas it appears in multiple windows and is encoded in multiple contexts for an overlapping split. Thus, improvements due to token fusion demonstrate the benefit of encoding representations of time points in diverse contexts. Fourth, enabling cross attention between base tokens in each window and neighboring fringe tokens increases performance, indicating the importance of this cross-attention mechanism for integration of contextual representations across neighboring windows. Lastly, we find that cross-window regularization that aligns window-specific $CLS$ tokens contributes notably to model performance. This result indicates that the model benefits from coherence of representations in $CLS$ tokens that are averaged across windows to implement classification. Overall, we find that the BolT model including all of its design elements yields the highest performance among all variants.}

\revhl{Compared to the non-windowed variant with $CLS$, the ablated variant with windowing and local $CLS$ tokens yields a notable performance improvement on the AAL atlas, albeit it shows relatively stable performance on the Schaefer atlas (Table \ref{tab:ablation}). Since the measured fMRI data and modeling procedures were identical for the two atlases, this difference is best attributed to an interaction between window size and the temporal characteristics of ROI responses. In theory, a larger $W$ can be suited to analyze relatively slow varying responses, and a smaller W can be suited to analyze relatively fast varying responses. A power spectral density analysis shows that ROI responses based on the AAL atlas that groups voxels based on anatomical proximity carry higher energy at high temporal frequencies, whereas responses based on the Schaefer atlas that groups voxels based on functional similarity carry higher energy at low temporal frequencies (not reported). In turn, we observe that the performance benefits from split time windows are maximized at $W$=20 for the AAL atlas, and at $W$=200 for the Schaefer atlas (not reported). Note, however, that BolT does not only use basic windowing as in the ablated variant, but it also leverages additional window-related design elements including cross-attention between base and fringe tokens, token fusion and cross-window regularization. These elements promote information exchange across separate time windows, increasing the effective temporal receptive field for each window. As such, a small W can serve to sensitively analyze a broad range of responses. Table \ref{tab:windowAblation} lists performance metrics for BolT under varying window sizes. We find that optimal or near-optimal performance is attained at a compact window size of $W$=20 commonly for both atlases. This result suggests that window-related design elements in BolT collectively introduce a degree of reliability against varying temporal frequency characteristics of BOLD responses.}

\begin{figure}[t]
  \centering
\includegraphics[width=\columnwidth]{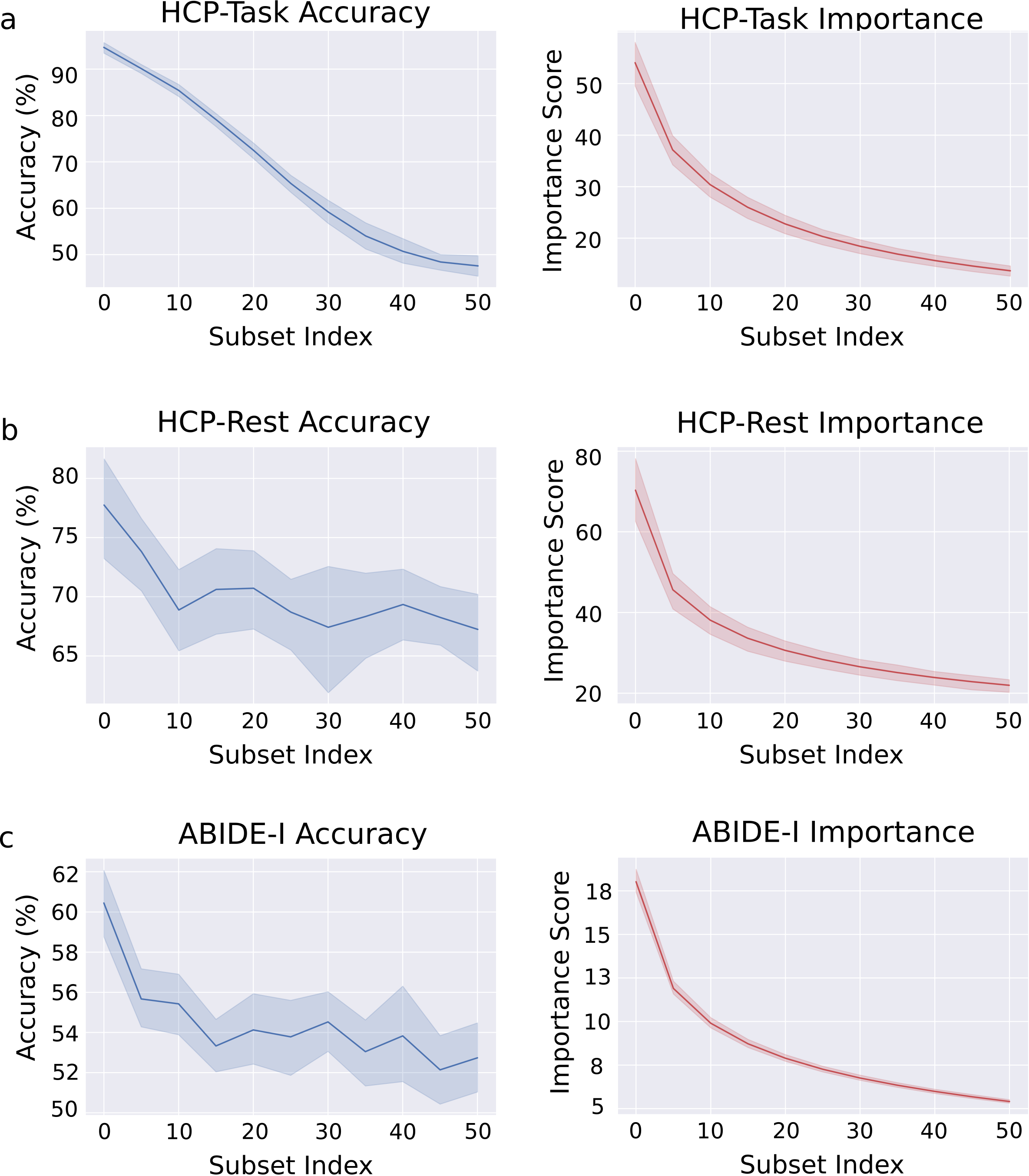}  
\caption{Accuracy of logistic-regression models receiving as input a subset of five important tokens, while the total importance of the subset is systematically varied. Tokens were ordered according to their importance weights, different subsets of five consecutive tokens in the ordered list were selected. Results are shown for varying subset index for (a) HCP-Task, (b) HCP-Rest, (c) ABIDE-I. Subset index refers to the offset within the ordered list for the selected subset. Importance score was taken as the average importance weight of tokens in a given subset normalized by the minimum importance weight within the time series. As such, total importance of the subset and detection accuracy show a general decrease with increasing subset index.}
  \label{fig:tokens}
  
\end{figure}

Next, we examined the efficacy of the explanatory technique that computes an importance weight for each BOLD token (see Fig. \ref{fig:relevancyFigure}). This importance weight is supposed to reflect the degree of discriminative information captured by the token for the respective detection task. We reasoned that if the explanatory technique computes reasonable weights, significant detection should be possible based on a subset of highly important tokens. To test this prediction, BOLD tokens in the time series were ordered according to their importance weights. ROI definitions based on the Schaefer atlas were used for this analysis, since they yielded better performance in BolT. Logistic-regression models were then built for the same detection task given as input a subset of five consecutive tokens \cite{tagliazucchi2012criticality,tagliazucchi2011spontaneous}. Table \ref{logisticTable} lists detection performance based on the most important subset of tokens against that based on a subset of randomly selected tokens. While random tokens perform near chance level, important tokens achieve substantially higher performance. We also reasoned that detection performance should scale with the overall importance of the selected token subset. To examine this issue, separate logistic-regression models were built while the overall importance of the token subset was systematically reduced (Fig. \ref{fig:tokens}). Detection performance elevates near-monotonically with increasing levels of token importance. Taken together, these results indicate that the importance weights returned by the explanatory technique closely reflect the contribution of individual tokens to model decisions.

\begin{table*}[tp]
  \caption{Performance of competing methods and BolT \revhl{using Schaefer atlas} for gender detection on the HCP-Rest dataset, task detection on the HCP-Task dataset and disease detection on the ABIDE-I dataset. Metrics are reported as mean$\pm$std across test folds. Boldface indicates the top-performing model in terms of each metric in individual classification tasks.}
  \label{tab:comparison_schaefer}
\centering
\resizebox{0.95\textwidth}{!}{%
\begin{NiceTabular}{ccccc|cccc|cccc}
\toprule
\thead{ } & \multicolumn{4}{c}{\thead{HCP-Rest \\ (Schaefer)}} &
\multicolumn{4}{c}{\thead{HCP-Task \\ (Schaefer)}} &
\multicolumn{4}{c}{\thead{ABIDE-I \\ (Schaefer)}} \\
\bottomrule
& \thead{Acc.(\%)} & \thead{Rec.(\%)} & \thead{Prec.(\%)} & \thead{AUC(\%)} & \thead{Acc.(\%)} & \thead{Rec.(\%)} & \thead{Prec.(\%)} & \thead{AUC(\%)}
& \thead{Acc.(\%)} & \thead{Rec.(\%)} & \thead{Prec.(\%)} & \thead{AUC(\%)}\\
\toprule
\thead{SVM} & \thead{75.92 \\ $\pm$4.78} & \thead{65.13 \\ $\pm$7.41} & \thead{78.63 \\ $\pm$6.27} & \thead{85.34 \\ $\pm$4.59} & \thead{95.57 \\ $\pm$0.87} & \thead{95.58 \\ $\pm$0.88} & \thead{95.61 \\ $\pm$0.86} & \thead{99.78 \\ $\pm$0.07} & \thead{65.72 \\ $\pm$4.11} & \thead{53.97 \\ $\pm$7.33} & \thead{66.26 \\ $\pm$6.68} & \thead{71.33 \\ $\pm$4.62}  \\
\toprule
\thead{BrainNetCNN} & \thead{84.16 \\ $\pm$3.73} & \thead{82.97 \\ $\pm$6.63} & \thead{82.39 \\ $\pm$3.04} & \thead{90.95 \\ $\pm$3.85} & \thead{96.42 \\$\pm$0.82} & \thead{96.42 \\ $\pm$0.87} & \thead{96.52 \\ $\pm$0.82} & \thead{99.81 \\$\pm$0.06} & \thead{66.78 \\ $\pm$3.90} & \thead{63.82 \\ $\pm$8.73} & \thead{65.50 \\ $\pm$4.63} & \thead{73.15 \\ $\pm$5.62 }   \\
\toprule
\thead{BrainGNN} & \thead{79.14 \\ $\pm$3.90} & \thead{78.96 \\ $\pm$5.43} & \thead{76.89 \\ $\pm$6.19} & \thead{86.31 \\ $\pm$2.51} & \thead{93.10 \\ $\pm$1.40} & \thead{93.11 \\ $\pm$1.43} & \thead{93.80 \\ $\pm$1.07} & \thead{99.67 \\ $\pm$0.12} & \thead{59.87 \\ $\pm$5.80} & \thead{47.86 \\ $\pm$12.21} & \thead{57.79 \\ $\pm$6.50} & \thead{63.86 \\ $\pm$6.63}  \\
\toprule
\thead{STAGIN} & \thead{82.51 \\ $\pm$3.69} & \thead{84.16 \\ $\pm$5.61} & \thead{79.20 \\ $\pm$4.71} & \thead{87.63 \\ $\pm$3.53} & \thead{99.27 \\ $\pm$0.43} & \thead{99.26 \\ $\pm$0.43} & \thead{99.27 \\ $\pm$0.42} & \thead{99.98 \\ $\pm$0.01} & \thead{61.70 \\ $\pm$3.56} & \thead{41.69 \\ $\pm$7.37} & \thead{63.63 \\ $\pm$7.03} & \thead{64.71 \\ $\pm$6.02}  \\
\toprule
\thead{LSTM} & \thead{81.59 \\ $\pm$4.03} & \thead{82.16 \\ $\pm$3.00} & \thead{78.99 \\ $\pm$6.37} & \thead{90.46 \\ $\pm$2.37} & \thead{98.35 \\ $\pm$0.65} & \thead{98.34 \\ $\pm$0.65} & \thead{98.41 \\ $\pm$0.58} & \thead{99.94 \\ $\pm$0.04} & \thead{64.55 \\ $\pm$5.41} & \thead{57.14 \\ $\pm$16.20} & \thead{62.85 \\ $\pm$6.98} & \thead{68.88 \\ $\pm$5.87}  \\
\toprule
\thead{CNN-LSTM} & \thead{80.77 \\ $\pm$3.83} & \thead{79.57 \\ $\pm$9.34} & \thead{80.38 \\ $\pm$9.27} & \thead{88.47 \\ $\pm$3.39} & \thead{99.04 \\ $\pm$0.59} & \thead{99.04 \\ $\pm$0.59} & \thead{99.07 \\ 0.56} & \thead{99.96 \\ $\pm$0.05} & \thead{65.49 \\ $\pm$5.70} & \thead{57.31 \\ $\pm$9.40} & \thead{64.79 \\ $\pm$7.57} & \thead{71.40 \\ $\pm$5.41}  \\
\toprule
\thead{GC-LSTM} & \thead{83.99 \\ $\pm$4.85} & \thead{76.90 \\ $\pm$16.09} & \thead{87.64 \\ $\pm$5.14} & \thead{93.85 \\ $\pm$1.35} & \thead{98.16 \\ $\pm$0.59} & \thead{98.16 \\ $\pm$0.59} & \thead{98.20 \\ $\pm$0.59} & \thead{99.96 \\ $\pm$0.02} & \thead{62.77 \\ $\pm$5.46} & \thead{60.24 \\ $\pm$19.56} & \thead{62.11 \\ $\pm$10.96} & \thead{68.28 \\ $\pm$4.84}  \\
\toprule
\thead{SwinT} & \thead{79.41 \\ $\pm$2.49} & \thead{77.35 \\ $\pm$4.54} & \thead{77.67 \\ $\pm$3.63} & \thead{87.49 \\ $\pm$1.71} & \thead{99.54 \\ $\pm$0.27} & \thead{99.54 \\ $\pm$0.27} & \thead{99.55 \\ $\pm$0.27} & \thead{\textbf{99.99} \\ \textbf{$\pm$0.00}} & \thead{68.56 \\ $\pm$4.74} & \thead{60.75 \\ $\pm$7.60} & \thead{68.27 \\ $\pm$6.56} & \thead{74.13 \\ $\pm$4.15}  \\
\toprule
\thead{Longformer} & \thead{83.24 \\ $\pm$3.34} & \thead{79.15 \\ $\pm$8.49} & \thead{84.03 \\ $\pm$5.45} & \thead{92.41 \\ $\pm$2.81} & \thead{99.46 \\ $\pm$0.39} & \thead{99.46 \\ $\pm$0.38} & \thead{99.47 \\ $\pm$0.38} & \thead{\textbf{99.99} \\ \textbf{$\pm$0.00}} & \thead{67.85 \\ $\pm$4.58} & \thead{57.15 \\ $\pm$15.79} & \thead{70.09 \\ $\pm$7.13} & \thead{74.87 \\ $\pm$3.99}  \\
\toprule
\thead{BaND} & \thead{83.61 \\ $\pm$4.03} & \thead{84.54 \\ $\pm$6.41} & \thead{80.87 \\ $\pm$5.42} & \thead{92.62 \\ $\pm$2.20} & \thead{99.24 \\ $\pm$0.46} & \thead{99.24 \\ $\pm$0.46} & \thead{99.26 \\ $\pm$0.43} & \thead{99.98 \\ $\pm$0.01} & \thead{65.48 \\ $\pm$3.04} & \thead{58.07 \\ $\pm$7.93} & \thead{64.47 \\ $\pm$4.76} & \thead{72.10 \\ $\pm$4.38} \\
\toprule
\thead{TFF} & \thead{87.36 \\ $\pm$3.68} & \thead{84.17 \\ $\pm$4.21} & \thead{87.89 \\ $\pm$5.09} & \thead{94.79 \\ $\pm$2.39} & \thead{99.08 \\ $\pm$0.25} & \thead{99.08 \\ $\pm$0.25} & \thead{99.11 \\ $\pm$0.24} & \thead{99.98 \\ $\pm$0.02} & \thead{66.73 \\ $\pm$5.33} & \thead{42.71 \\ $\pm$16.49} & \thead{\textbf{78.93} \\ \textbf{$\pm$10.19}} & \thead{75.44 \\ $\pm$3.94}  \\
\toprule
\thead{IFT-Net} & \thead{82.97 \\ $\pm$2.84} & \thead{79.76 \\ $\pm$10.28} & \thead{83.72 \\ $\pm$7.27} & \thead{93.12 \\ $\pm$2.05} & \thead{97.58 \\ $\pm$3.62} & \thead{97.58 \\ $\pm$3.62} & \thead{98.08 \\ $\pm$2.59} & \thead{99.91 \\ $\pm$0.16} & \thead{61.88 \\ $\pm$5.52} & \thead{51.59 \\ $\pm$23.88} & \thead{63.82 \\ $\pm$9.42} & \thead{68.50 \\ $\pm$6.44}  \\
\toprule
\thead{HATNet} & \thead{85.72 \\ $\pm$2.64} & \thead{84.38 \\ $\pm$5.89} & \thead{84.57 \\ $\pm$3.68} & \thead{93.66 \\ $\pm$2.26} & \thead{99.37 \\ $\pm$0.33} & \thead{99.36 \\ $\pm$0.33} & \thead{99.38 \\ $\pm$0.32} & \thead{\textbf{99.99} \\ \textbf{$\pm$0.00}} & \thead{64.66 \\ $\pm$4.53} & \thead{57.67 \\ $\pm$6.88} & \thead{62.99 \\ $\pm$5.76} & \thead{68.92 \\ $\pm$4.95}  \\
\toprule
\thead{BolT} & \thead{\textbf{91.85} \\ \textbf{$\pm$3.05}} & \thead{\textbf{90.58} \\ \textbf{$\pm$4.97}} & \thead{\textbf{91.51} \\ \textbf{$\pm$3.07}} & \thead{\textbf{97.35} \\ \textbf{$\pm$1.06}} & \thead{\textbf{99.66} \\ \textbf{$\pm$0.35}} & \thead{\textbf{99.66} \\ \textbf{$\pm$0.35}} & \thead{\textbf{99.67} \\ \textbf{$\pm$0.34}} & \thead{\textbf{99.99} \\ \textbf{$\pm$0.00}} & \thead{\textbf{71.28} \\ \textbf{$\pm$4.62}} & \thead{\textbf{64.85} \\ \textbf{$\pm$7.94}} & \thead{71.32 \\ $\pm$7.35} & \thead{\textbf{77.56} \\ \textbf{$\pm$3.44}}  \\
\toprule

\end{NiceTabular}}
\end{table*}

\begin{table*}[tp]
  \caption{Performance of competing methods and BolT \revhl{using AAL atlas} for gender detection on the HCP-Rest dataset, task detection on the HCP-Task dataset and disease detection on the ABIDE-I dataset. Metrics are reported as mean$\pm$std across test folds. Boldface indicates the top-performing model in terms of each metric in individual classification tasks.}
  \label{tab:comparison_aal}
\centering
\resizebox{0.95\textwidth}{!}{%
\begin{NiceTabular}{ccccc|cccc|cccc}
\toprule
\thead{ } & \multicolumn{4}{c}{\thead{HCP-Rest \\ (AAL)}} &
\multicolumn{4}{c}{\thead{HCP-Task \\ (AAL)}} &
\multicolumn{4}{c}{\thead{ABIDE-I \\ (AAL)}} \\
\bottomrule
& \thead{Acc.(\%)} & \thead{Rec.(\%)} & \thead{Prec.(\%)} & \thead{AUC(\%)} & \thead{Acc.(\%)} & \thead{Rec.(\%)} & \thead{Prec.(\%)} & \thead{AUC (\%)}
& \thead{Acc.(\%)} & \thead{Rec.(\%)} & \thead{Prec.(\%)} & \thead{AUC(\%)}\\
\toprule
\thead{SVM} & \thead{71.71 \\ $\pm$3.76} & \thead{65.40 \\ $\pm$11.86} & \thead{70.55 \\ $\pm$3.28} & \thead{ 78.30 \\ $\pm$4.09} & \thead{88.11 \\ $\pm$1.46} & \thead{88.14 \\ $\pm$1.46} & \thead{88.25 \\ $\pm$1.38} & \thead{98.54 \\ $\pm$0.31} & \thead{65.60 \\ $\pm$3.14} & \thead{55.00 \\ $\pm$7.40} & \thead{64.10 \\ $\pm$5.77} & \thead{71.16 \\ $\pm$4.18} \\
\toprule
\thead{BrainNetCNN} & \thead{71.16 \\ $\pm$4.91} & \thead{67.40 \\ $\pm$9.21} & \thead{69.20 \\ $\pm$6.47} & \thead{78.99 \\ $\pm$5.12 } & \thead{90.56 \\ $\pm$1.06} & \thead{90.58 \\ $\pm$1.07} & \thead{90.73 \\ $\pm$1.04} & \thead{99.04 \\ $\pm$0.27} & \thead{64.01 \\ $\pm$6.04}  & \thead{60.24 \\ $\pm$9.33}  & \thead{61.86 \\ $\pm$ 6.75}  & \thead{69.79 \\ $\pm$6.16}   \\
\toprule
\thead{BrainGNN} & \thead{69.79 \\ $\pm$4.59} & \thead{72.40 \\ $\pm$9.70} & \thead{65.25 \\ $\pm$4.36} & \thead{77.85 \\ $\pm$4.59} & \thead{79.73 \\ $\pm$5.92} & \thead{79.85 \\ $\pm$5.88} & \thead{81.72 \\ $\pm$4.77} & \thead{97.05 \\ $\pm$1.29} & \thead{61.40 \\ $\pm$4.66} & \thead{58.30 \\ $\pm$10.86} & \thead{58.40 \\ $\pm$6.74} & \thead{65.68 \\ $\pm$5.61}  \\
\toprule
\thead{STAGIN} & \thead{76.18 \\ $\pm$3.10} & \thead{72.20 \\ $\pm$9.22} & \thead{75.00 \\ $\pm$3.75} & \thead{83.07 \\ $\pm$3.15} & \thead{98.87 \\ $\pm$0.60} & \thead{98.86 \\ $\pm$0.61} & \thead{98.91 \\ $\pm$0.57} & \thead{99.95 \\ $\pm$0.03} & \thead{61.52 \\ $\pm$3.49} & \thead{52.69 \\ $\pm$7.38} & \thead{60.12 \\ $\pm$4.94} & \thead{66.68 \\ $\pm$4.36}  \\
\toprule
\thead{LSTM} & \thead{73.25 \\ $\pm$4.48} & \thead{67.19 \\ $\pm$8.44} & \thead{73.10 \\ $\pm$7.59} & \thead{81.96 \\ $\pm$2.90} & \thead{96.96 \\ $\pm$0.69} & \thead{96.97 \\ $\pm$0.69} & \thead{97.06 \\ $\pm$0.64} & \thead{99.88 \\ $\pm$0.06} & \thead{63.06 \\ $\pm$3.96} & \thead{45.81 \\ $\pm$22.76} & \thead{64.05 \\ $\pm$13.93} & \thead{70.25 \\ $\pm$3.85}  \\
\toprule
\thead{CNN-LSTM} & \thead{74.81 \\ $\pm$3.15} & \thead{66.40 \\ $\pm$9.20} & \thead{76.25 \\ $\pm$5.28} & \thead{82.88 \\ $\pm$3.30} & \thead{97.77 \\ $\pm$0.60} & \thead{97.76 \\ $\pm$0.61} & \thead{97.83 \\ $\pm$0.57} & \thead{99.91 \\ $\pm$0.05} & \thead{63.65 \\ $\pm$5.42} & \thead{49.47 \\ $\pm$17.04} & \thead{66.49 \\ $\pm$11.86} & \thead{68.78 \\ $\pm$4.13}  \\
\toprule
\thead{GC-LSTM} & \thead{77.27 \\ $\pm$5.29} & \thead{77.79 \\ $\pm$14.57} & \thead{77.61 \\ $\pm$11.50} & \thead{89.59 \\ $\pm$1.93} & \thead{93.14 \\ $\pm$1.39} & \thead{93.14 \\ $\pm$1.39} & \thead{93.74 \\ $\pm$1.09} & \thead{99.57 \\ $\pm$0.14} & \thead{60.05 \\ $\pm$5.36} & \thead{48.27 \\ $\pm$19.80} & \thead{61.04 \\ $\pm$9.04} & \thead{64.67 \\ $\pm$6.90}  \\
\toprule
\thead{SwinT} & \thead{78.37 \\ $\pm$4.00} & \thead{76.60 \\ $\pm$6.69} & \thead{76.14 \\ $\pm$4.07} & \thead{84.84 \\ $\pm$3.58} & \thead{99.23 \\ $\pm$0.38} & \thead{99.23 \\ $\pm$0.38} & \thead{99.25 \\ $\pm$0.36} & \thead{\textbf{99.99} \\ \textbf{$\pm$0.01}} & \thead{66.78 \\ $\pm$4.25} & \thead{\textbf{60.48} \\ $\pm$\textbf{6.38}} & \thead{65.56 \\ $\pm$5.57} & \thead{72.92 \\ $\pm$4.30}  \\
\toprule
\thead{Longformer} & \thead{76.28 \\ $\pm$4.30} & \thead{63.00 \\ $\pm$13.94} & \thead{83.53 \\ $\pm$9.68} & \thead{87.80 \\ $\pm$3.10} & \thead{99.11 \\ $\pm$0.42} & \thead{99.11 \\ $\pm$0.43} & \thead{99.13 \\ $\pm$0.41} & \thead{\textbf{99.99} \\ \textbf{$\pm$0.00}} & \thead{64.77 \\ $\pm$4.71} & \thead{58.69 \\ $\pm$18.18} & \thead{64.99 \\ $\pm$9.42} & \thead{71.42 \\ $\pm$4.70}  \\
\toprule
\thead{BaND} & \thead{78.55 \\ $\pm$4.12} & \thead{70.40 \\ $\pm$8.38} & \thead{80.46 \\ $\pm$5.18} & \thead{87.39 \\ $\pm$2.84} & \thead{98.16 \\ $\pm$0.45} & \thead{98.16 \\ $\pm$0.45} & \thead{98.20 \\ $\pm$0.41} & \thead{99.93 \\ $\pm$0.03} & \thead{63.12 \\ $\pm$3.60} & \thead{45.18 \\ $\pm$12.53} & \thead{65.25 \\ $\pm$6.33} & \thead{68.65 \\ $\pm$3.61}  \\
\toprule
\thead{TFF} & \thead{82.57 \\ $\pm$4.05} & \thead{81.80 \\ $\pm$8.59} & \thead{81.34 \\ $\pm$7.41} & \thead{91.14 \\ $\pm$2.98} & \thead{97.43 \\ $\pm$1.00} & \thead{97.43 \\ $\pm$1.00} & \thead{97.55 \\ $\pm$0.90} & \thead{99.90 \\ $\pm$0.07} & \thead{65.84 \\ $\pm$3.87} & \thead{46.30 \\ $\pm$15.83}  & \thead{\textbf{73.30} \\ \textbf{$\pm$8.86}} & \thead{74.14 \\ $\pm$4.16}  \\
\toprule
\thead{IFT-Net} & \thead{77.72 \\ $\pm$4.85} & \thead{76.00 \\ $\pm$11.76} & \thead{76.57 \\ $\pm$7.28} & \thead{86.69 \\ $\pm$2.62} & \thead{95.22 \\ $\pm$5.58} & \thead{95.17 \\ $\pm$5.67} & \thead{96.19 \\ $\pm$3.80} & \thead{99.87 \\ $\pm$0.19} & \thead{58.26 \\ $\pm$4.72} & \thead{32.11 \\ $\pm$24.85} & \thead{54.23 \\ $\pm$25.90} & \thead{64.99 \\ $\pm$4.64} \\
\toprule
\thead{HATNet} & \thead{78.37 \\ $\pm$2.15} & \thead{74.60 \\ $\pm$5.58} & \thead{77.23 \\ $\pm$1.54} & \thead{87.43 \\$\pm$2.96} & \thead{97.97 \\ $\pm$0.35} & \thead{97.96 \\ $\pm$0.35} & \thead{98.02 \\ $\pm$0.32} & \thead{99.96 \\ $\pm$0.02} &  \thead{61.16 \\ $\pm$5.64} & \thead{54.59 \\ $\pm$7.15} & \thead{58.79 \\ $\pm$6.55} & \thead{66.52 \\ $\pm$5.10}  \\
\toprule
\thead{BolT} & \thead{\textbf{87.31} \\ \textbf{$\pm$2.69}} & \thead{\textbf{86.99} \\ \textbf{$\pm$4.49}} & \thead{\textbf{85.65} \\ \textbf{$\pm$4.01}} & \thead{\textbf{94.29} \\ \textbf{$\pm$2.05}} & \thead{\textbf{99.52} \\ \textbf{$\pm$0.39}} & \thead{\textbf{99.52} \\ \textbf{$\pm$0.40}} & \thead{\textbf{99.54} \\ \textbf{$\pm$0.38}} & \thead{\textbf{99.99} \\ \textbf{$\pm$0.00}} & \thead{\textbf{68.14} \\ \textbf{$\pm$2.81}} & \thead{\textbf{60.49} \\ \textbf{$\pm$4.22}} & \thead{67.52 \\ $\pm$4.07} & \thead{\textbf{74.30} \\ \textbf{$\pm$3.69}}  \\
\toprule

\end{NiceTabular}}
\end{table*}

\subsection{Comparative demonstration of BolT}
We demonstrated BolT for three main tasks in fMRI analysis: gender detection on HCP-Rest, cognitive task detection on HCP-Task, and disease detection on ABIDE-I datasets. BolT was demonstrated against state-of-the-art \revhl{traditional (SVM), CNN (BrainNetCNN), GNN (BrainGNN, STAGIN), RNN (LSTM, CNN-LSTM, GC-LSTM), and
 transformer (SwinT, Longformer, BaND, TFF, IFT-Net, HATNet) baselines.} Demonstrations were performed using ROI definitions extracted via two different brain atlases. \revhl{Performance metrics for competing methods for Schaefer atlas are listed in Table \ref{tab:comparison_schaefer}, and those for AAL atlas are listed in Table \ref{tab:comparison_aal}.} For each detection task and based on each atlas, BolT outperforms all competing methods in each metric (p$<$0.05, Wilcoxon signed-rank test), \revhl{except for TFF that offers higher precision on ABIDE-I, SwinT that offers similar recall on ABIDE-I (AAL atlas), and SwinT, Longformer, and HATNet that offer similar AUC on HCP-Task. On average across atlases in gender detection, BolT improves (accuracy, recall, precision, AUC) by (8.39, 11.13, 7.41, 5.87)\% over transformer baselines, (10.96, 13.78, 9.58, 7.95)\% over RNN baselines, (12.67, 11.85, 14.49, 12.10)\% over GNN baselines, (11.92, 13.59, 12.78, 10.84)\% over the CNN baseline, and (15.76, 23.51, 13.99, 14.00)\% over the traditional baseline. In cognitive task detection, BolT achieves improvements of (1.14, 1.14, 1.00, 0.03)\% over transformer baselines, (2.35, 2.35, 2.21, 0.12)\% over RNN baselines, (6.84, 6.82, 6.18, 0.82)\% over GNN baselines, (6.09, 6.09, 5.98, 0.56)\% over the CNN baseline, and (7.75, 7.73, 7.67, 0.82)\% over the traditional baseline. Finally, in disease detection, BolT achieves improvements of (5.11, 10.56, -, 4.88)\% over transformer baselines, (6.44, 9.63, 5.86, 7.22)\% over RNN baselines, (8.58, 12.53, 9.43, 10.69)\% over GNN baselines, (4.31, 0.64, 5.73, 4.46)\% over the CNN baseline, and (4.05, 8.15, 4.23, 4.68)\% over the traditional baseline.} Taken together, these results indicate that BolT enables significant performance benefits in detection tasks over prior traditional and DL methods. 
 
 \revhl{In general, we observe that DL models yield superior performance to the traditional SVM baseline on HCP-Rest and HCP-Task, whereas SVM outperforms CNN, GNN, RNN, and a subset of transformer baselines in disease detection on ABIDE-I. Note that HCP-Rest and HCP-Task were acquired using relatively standardized protocols and scanner hardware in a compact set of imaging sites. In contrast, ABIDE-I was curated by aggregating data from a larger number of sites with more substantial variations in imaging protocols and hardware. In turn, the resultant data heterogeneity can limit generalization performance for DL methods with relatively high complexity, while the simpler SVM method starts performing competitively. That said, we observe that windowed transformer models including BolT still outperform SVM in this case, implying a degree of reliability against data heterogeneity due to the generalization capabilities of self-attention operators combined with local sensitivity from split time windows. We also observe that all competing methods yield notably higher performance on HCP-Task, compared to HCP-Rest and ABIDE-I. This is expected as detecting divergent cognitive tasks from task-based fMRI scans that elicit responses in largely non-overlapping brain networks is relatively easier compared to detection tasks on resting-state fMRI scans. Here, we preferred to report HCP-Task since it is a highly relevant, benchmark dataset that is frequently reported in methodology studies on task-based fMRI analysis. Yet, future studies are warranted to examine the utility of BolT in detecting BOLD-response differences among more similar cognitive tasks driving partly overlapping brain networks.}


\begin{figure*}[t]
  \centering
\includegraphics[width=0.8\linewidth]{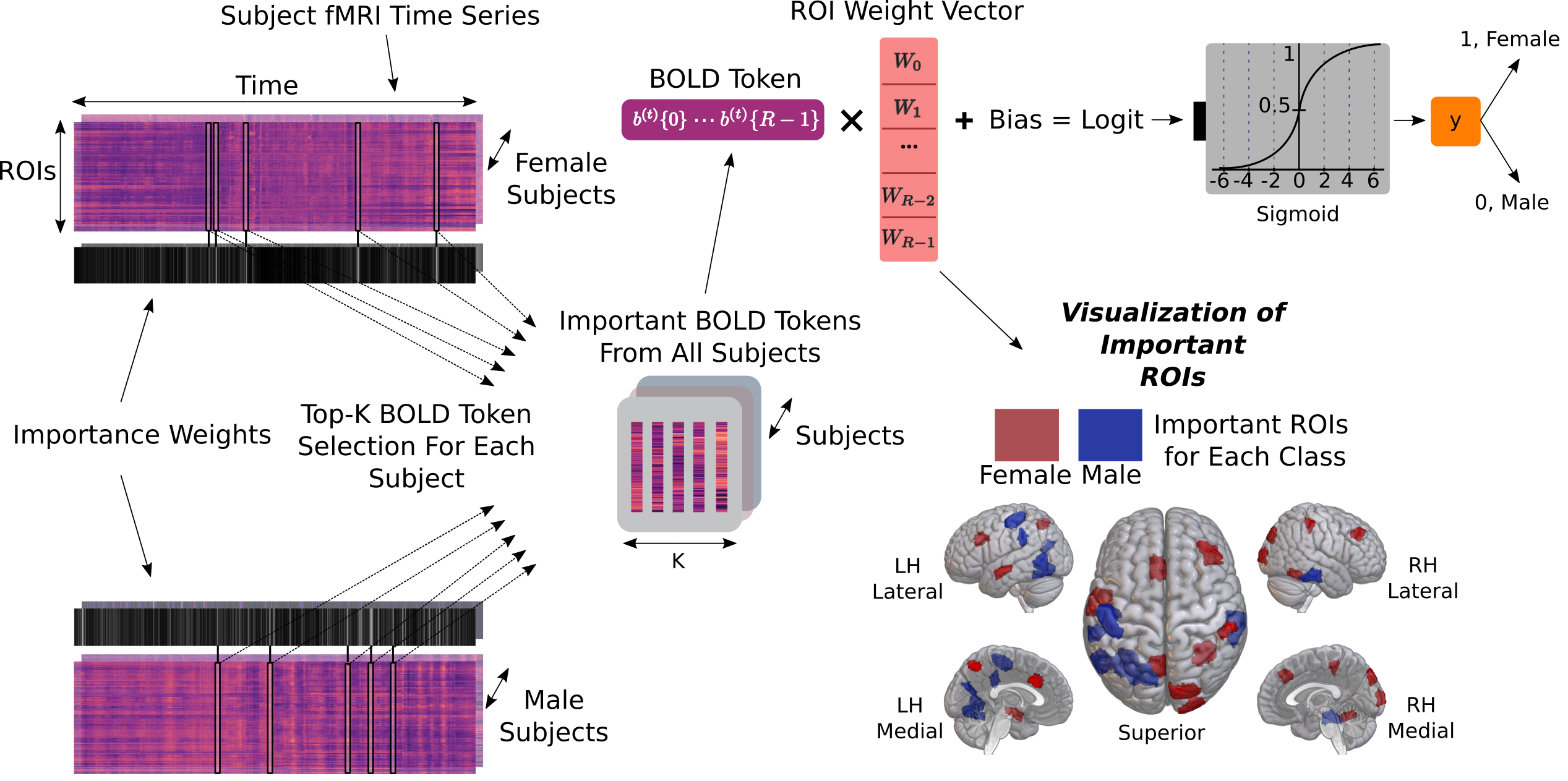}  
  \caption{Landmark time points (i.e. BOLD tokens) selected by BolT were used to identify brain regions critical for gender detection in HCP-Rest. A collection of K=5 tokens were retrieved from each subject, characterizing responses across R ROIs. Next, a logistic regression model was trained to map the tokens in landmark time points onto the associated output class. Model weights reflect each ROI's contribution to the classification decision. For each class, the top 2 percent of most influential ROIs were visualized (i.e. female in red color, male in blue color).}
  \label{fig:brainMapp}
\end{figure*}

\subsection{Explanability of BolT}
To interpret the spatio-temporal patterns of brain activation that contribute to BolT's decisions, we employed the explanatory technique to calculate token importance weights. Importance weights for each cognitive task in HCP-Task are shown in Fig. \ref{fig:relevancyFigure}. Landmark time points of high importance closely align with transitions in the temporal structure of task variables following an offset due to hemodynamic delay. For instance, periods of target maintenance following target appearance are attributed high importance in the working memory task, corresponding to abrupt changes in activation \cite{tagliazucchi2011spontaneous}.

\begin{figure}[t]
  \centering
\includegraphics[width=\columnwidth]{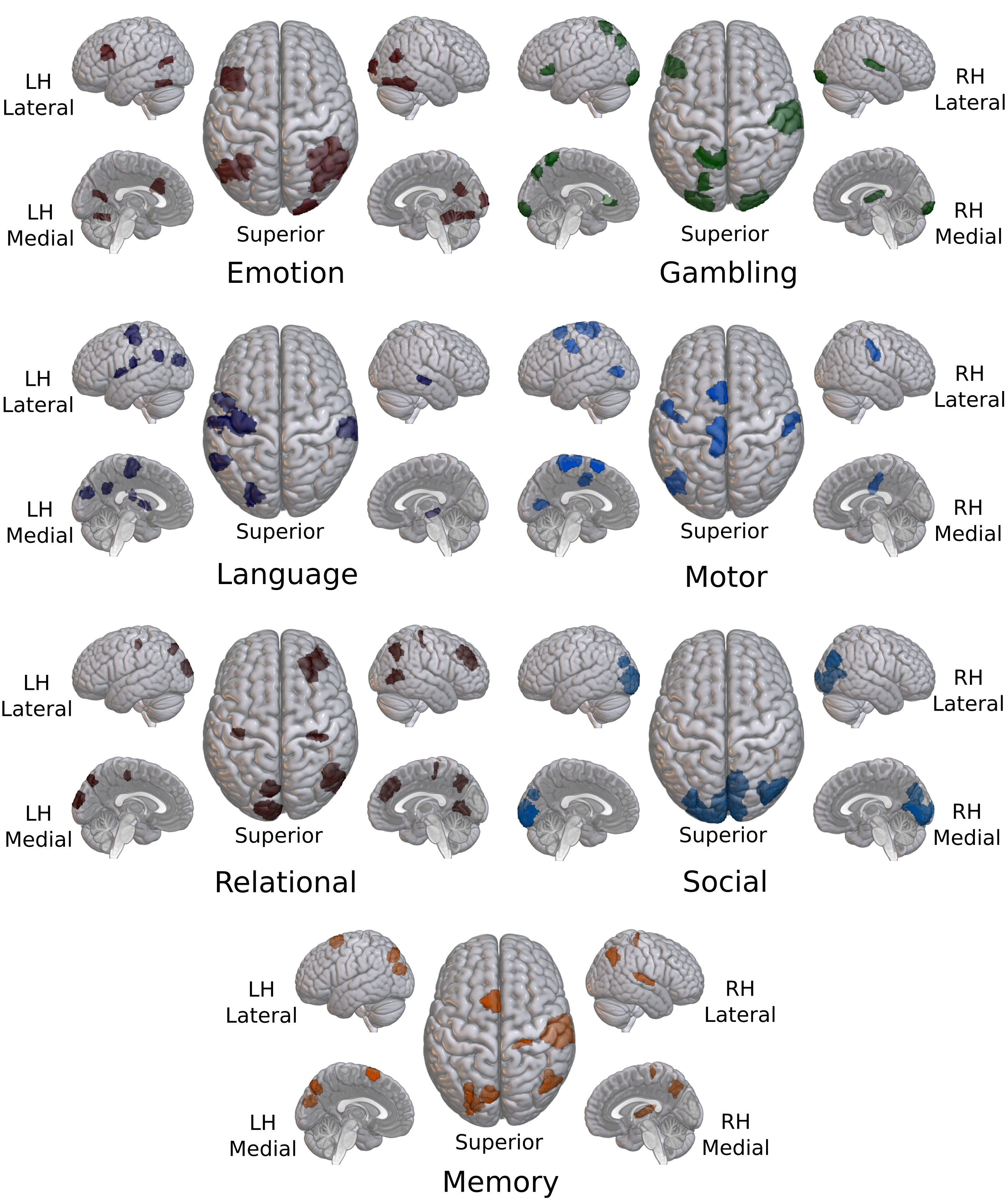}  
  \caption{Critical ROIs for cognitive task detection in HCP-Task. For each task, the top 2 percent of most influential ROIs were visualized. Elevated BOLD responses in highlighted ROIs imply the presence of the associated task.}
\label{fig:taskBrainMapping}
  
\end{figure}

\begin{figure}[t]
  \centering
    \includegraphics[width=0.8\linewidth]{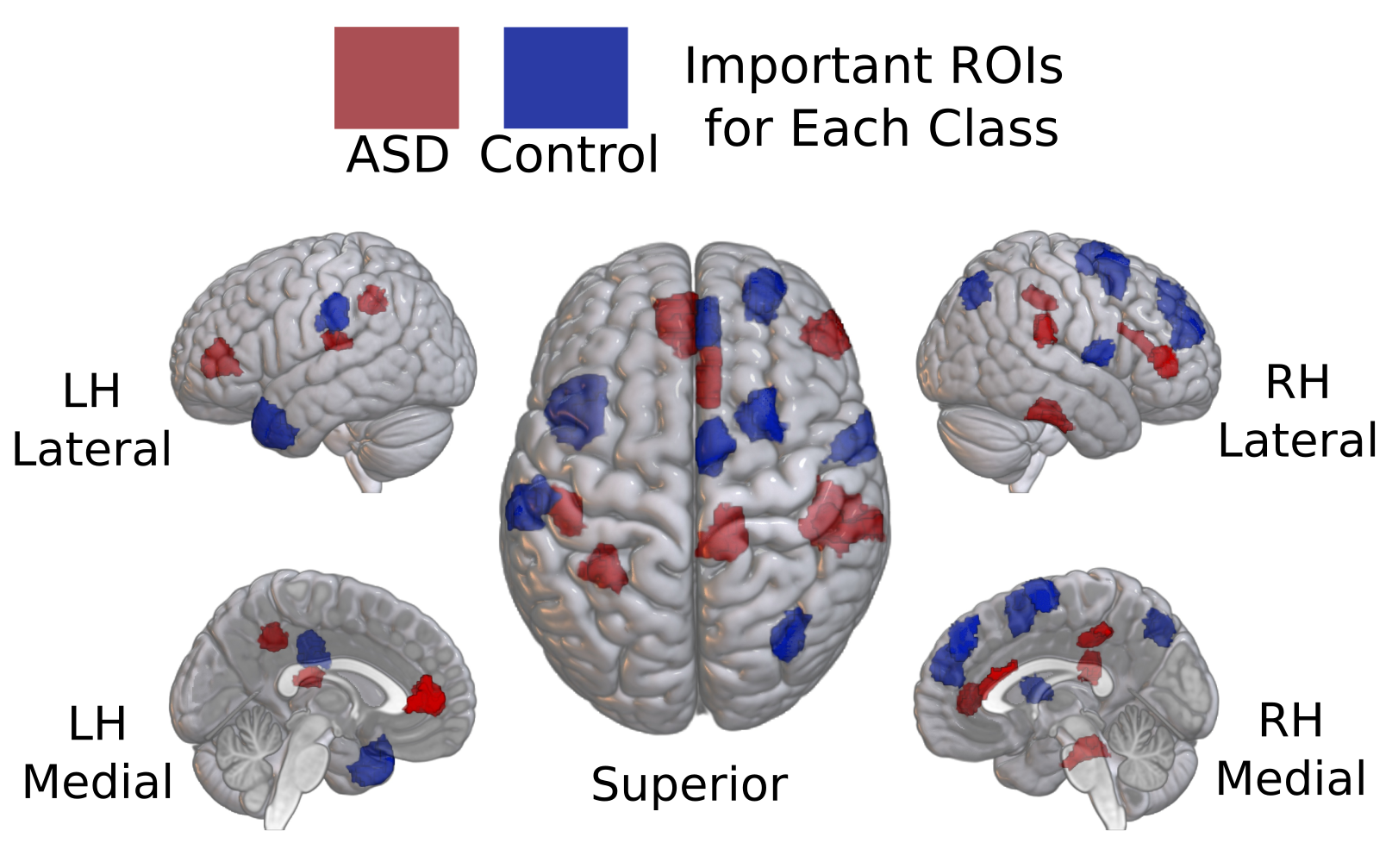}
  \caption{Critical ROIs for ASD detection in ABIDE-I. The top 2 percent of most influential ROIs were visualized for ASD patients and healthy controls. Elevated BOLD responses in red ROIs imply the presence of the ASD condition, whereas elevated responses in blue ROIs imply a healthy condition.}
\label{fig:autismBrainMapping}
  
\end{figure}

We then leveraged the landmark time points to identify brain regions critical for the detection tasks. To do this, a logistic-regression model was trained on top-five most important BOLD tokens, and model weights were taken to reflect the importance of individual ROIs for task performance \cite{rahman2022interpreting}. As shown in Fig. \ref{fig:brainMapp} for gender detection, we find important ROIs across the attention and somatosensory networks in male subjects, and ROIs in prefrontal/frontal cortices and default mode network (DMN) in female subjects. This is consistent with previous reports on stronger FC features across sensorimotor cortices in males and across DMN in females \cite{ritchie2018sex,filippi2013organization}. We further find important ROIs in visual networks for both genders. This result is aligned with a recent report suggesting that responses in visual regions might implicitly represent gender-discriminating information \cite{kim2021learning}. As shown in Fig. \ref{fig:taskBrainMapping} for task detection, brain regions implicated with the target task are attributed high importance (e.g., sensorimotor regions in the Motor task, temporal regions in the Language task). As shown in Fig. \ref{fig:autismBrainMapping} for ASD detection, we find important ROIs in healthy controls across the frontal-parietal network (FPN), thought to mediate goal-oriented, cognitively demanding behavior \cite{uddin2019towards}. In contrast, ASD patients manifest important ROIs across DMN, with commonly reported over-activation in ASD \cite{buckner2008brain,abraham2017deriving,chen2021excessive}. Taken together, these results indicate that BolT effectively captures task-relevant patterns of brain activation in both normal and disease states.

\section{Discussion}
Here, we introduced a transformer architecture that efficiently captures local-to-global representations of time series to perform detection tasks based on fMRI scans. The proposed architecture learns latent representations of fMRI data via a novel fused window attention mechanism that incorporates long-range context with linear complexity in terms of scan length. Detection is then performed based on learned high-level classification tokens regularized across time windows. Demonstrations were performed on resting-state and task-based fMRI data with superior performance against state-of-the-art baselines including convolutional, graph and transformer models. 

In this study, we primarily built classification models with categorical output variables for gender, cognitive task, and disease. To improve classification performance, learnable $CLS$ tokens were included that provide a condensed high-level representation of corresponding time windows. Note that the human brain does not only represent categorical variables, but it is also assumed to carry information regarding continuous stimulus or task features \cite{ccukur2013attention}. To analyze cortical representations of such continuous features, BolT can be adapted to instead build regression models \cite{nishimoto2011,vanrullen2019reconstructing,li2018brain}. To do this, latent representations of BOLD tokens in downstream layers of BolT can be coupled with linear or ReLU activation functions. This can enable BolT to decode continuous stimulus or task variables from fMRI scans. 

A common approach to interrogate brain function in computational neuroimaging studies is to build decoding models that predict stimulus or task features given as input measured brain activations \cite{laconte2011decoding,andersson2011real}. Following this framework, here we used BolT to build decoding models that detect external features based on measured BOLD responses. An alternative framework rests on analysis of brain function by building encoding models that predict brain activations given as input stimulus/task features \cite{nishimoto2021modeling,celik2021cortical,shahdloo2022task,anderson2016representational,ngo2022transformer}. In cognitive neuroimaging studies, the experimental time course for the stimulus and/or cognitive task can be taken as input to BolT, and voxel-wise regression models can be built to estimate measured BOLD responses. It remains important future work to assess the efficacy of BolT in training encoding models. 

Literature suggests that resting-state fMRI scans carry idiosyncratic information regarding disease progression in neurodevelopmental disorders \cite{uddin2010typical,hohenfeld2018resting}. Based on this literature, we considered ASD detection using solely information from resting-state fMRI scans. Recent studies suggest that auxiliary information on patient demographics or scan protocols might help facilitate disease detection \cite{dvornek2018combining}. Moreover, some neurological diseases such as Alzheimer's or dementia have complementary imaging signatures in other modalities such as structural or diffusion-weighted MRI \cite{roman2012contribution}. Thus, it is reasonable to expect that disease detection performance with BolT can be further improved by incorporating auxiliary information as well as additional imaging modalities. Auxiliary information can be integrated via bypass channels near the output layers of BolT, whereas additional imaging modalities can be incorporated as added input channels alongside fMRI data. 

As commonly practiced in many fMRI studies, here we first normalized each subject's brain volume onto an anatomical template, and then used an anatomical atlas to define brain ROIs. Average BOLD responses in individual ROIs were then provided as input to BolT. Note that this approach ensures relatively consistent and comprehensive ROI definitions across subjects, permitting analyses in brain regions that do not have well established functional-localization procedures \cite{flandin2002improved}. Yet, spatial registration to a common template involves a potentially lossy transformation of fMRI data. Such losses can be mitigated by defining ROIs in the brain spaces of individual subjects as opposed to a template. To do this, the registration transform between the subject and template brain spaces can be estimated. ROI boundaries in the template brain space can then be backprojected onto the individual subject brain space by inverting the estimated transformation \cite{shahdloo2020biased,shahdloo2022task}. Alternatively, a CNN model can also incorporated in BolT to perform spatial encoding of volumetric MRI data prior to processing with the transformer blocks \cite{nguyen2020attend,malkiel2021pre}.

Here we trained all competing models from scratch on fMRI data from several hundred subjects for each detection task. Given their relatively higher complexity against convolutional models, transformers are generally considered to require substantial datasets for successful learning \cite{dosovitskiy2020image,gungor2022transms}. In applications where only compact datasets are available, pre-training and transfer learning procedures can be adopted to initialize the network weights in transformer architectures \cite{devlin2018bert,dalmaz2021resvit}. Reliable augmentation via image synthesis based on advanced procedures such as diffusion modeling can also help alleviate data scarcity \cite{dar2022adaptive,ozbey2022unsupervised}. Alternatively, complexity of self-attention modules can be mitigated by replacing regular dot-product attention operators with efficient kernelized operators \cite{zhang2022diffusion}. Federated learning across multiple institutions might facilitate learning from large, diverse datasets without introducing privacy risks \cite{elmas2022federated,dalmaz2022one}. Lastly, unsupervised learning strategies can also be adopted to permit training on partially labeled fMRI datasets from a larger subject cohort \cite{malkiel2021pre,korkmaz2022unsupervised}. A systematic exploration of the data efficiency of BolT against competing models remains an important topic for future research. 


\section{Conclusion}
In this study, we introduced a novel transformer model to improve classification performance on fMRI time series. BolT leverages fused window attention to capture local interactions among temporally-overlapped time windows, and hierarchically grows window overlap to capture global representations. Token fusion and cross-window regularization are used to effectively integrate latent representations across the time series. Here, demonstrations were performed for gender and disease detection from resting-state fMRI and task detection from task-based fMRI. Furthermore, an explanatory technique was devised to interpret model decisions in terms of landmark time points and brain regions. Collectively, the proposed approach holds great promise for sensitive and explainable analysis of multi-variate fMRI data. BolT may help detect other neurological disorders with characteristic influences on fMRI activation patterns, and classification of more intricate task variables during cognitive processing. 



\bibliographystyle{TMI_STYLE/IEEEtran}
\bibliography{Papers}

\end{document}